\documentclass[a4paper,fleqn,usenatbib]{mnras}

\usepackage[T1]{fontenc}
\usepackage{ae,aecompl}

\usepackage{graphicx}	% Including figure files
\usepackage{amsmath}	% Advanced maths commands
\usepackage{amssymb}	% Extra maths symbols
\usepackage{txfonts}
\usepackage{color, colortbl}
\usepackage{enumitem}
\usepackage{yfonts}
\usepackage{ulem}

\newcommand{\be}{\begin{equation}}
\newcommand{\ee}{\end{equation}}

\def\rewmgii{\hbox{$W_\textnormal{r}^{\lambda 2796}$}}%

\title[A MEGAFLOW NOEMA pilot program]{MusE GAs FLOw and Wind (MEGAFLOW) VII. A NOEMA pilot program to probe molecular gas in galaxies with measured circumgalactic gas flows}

\author[J. Freundlich et al.]{Jonathan Freundlich,$^{1,2,3}$\thanks{E-mail: jonathan.freundlich@astro.unistra.fr}
Nicolas F. Bouch\'e,$^{4}$ 
Thierry Contini,$^{5}$
Emanuele Daddi,$^{6}$
\newauthor
Johannes Zabl,$^{4}$
 Ilane Schroetter,$^{7}$ 
Leindert Boogaard,$^{8}$ 
Johan Richard$^{4}$
\\
$^1$School of Physics and Astronomy, Tel Aviv University, Tel Aviv 69978, Israel\\
$^2$Centre for Astrophysics and Planetary Science, Racah Institute of Physics, The Hebrew University, Jerusalem 91904, Israel\\
$^3$Universit\'e de Strasbourg, CNRS UMR 7550, Observatoire astronomique de Strasbourg, 67000 Strasbourg, France\\
$^4$Univ Lyon, Univ Lyon1, Ens de Lyon, CNRS, Centre de Recherche Astrophysique de Lyon UMR5574, F-69230 Saint-Genis-Laval, France\\
$^5$Institut de Recherche en Astrophysique et Plan\'etologie (IRAP), Universit\'e de Toulouse, CNRS, UPS, F-31400 Toulouse, France\\
$^6$CEA, Irfu, DAp, AIM, Universit\'e Paris-Saclay, Universit\'e de Paris, CNRS, F-91191 Gif-sur-Yvette, France\\
$^7$GEPI, Observatoire de Paris, CNRS-UMR8111, PSL Research University, Univ. Paris Diderot, 5 place Jules Janssen, 92195 Meudon, France\\
$^8$Leiden Observatory, Leiden University, P.O. Box 9513, NL-2300 RA Leiden\\
}

\date{Accepted XXX. Received YYY; in original form ZZZ}

\pubyear{2020}

\begin{document}
\label{firstpage}
\pagerange{\pageref{firstpage}--\pageref{lastpage}}
\maketitle

\begin{abstract}
	
We present a pilot program using IRAM's NOrthern Extended Millimeter Array (NOEMA) to probe the molecular gas reservoirs of six $z=0.6-1.1$ star-forming galaxies whose circumgalactic medium has been observed in absorption along quasar lines-of-sight as part of the MusE GAs FLOw and Wind (MEGAFLOW) survey and for which we have estimates of either the accretion or the outflow rate. This program is motivated by testing the quasi equilibrium model and the compaction scenario describing the evolution of galaxies along the main sequence of star formation, which imply tight relations between the gas content, the star formation activity, and the amount of gas flowing in and out. We report individual carbon monoxide CO(4-3), CO(3-2) and dust continuum upper limits, as well as stacked CO detections over the whole sample and the three galaxies identified with outflows. The resulting molecular gas fractions and depletion times are compatible with published scaling relations established within a mass-selected sample, indicating that galaxies selected through their absorption follow similar relations on average. We further detect the dust continuum of three of the quasars and a strong emission line in one of them, which we identify as CO(4-3). Extending the sample to more galaxies and deeper observations will enable to quantify how the molecular gas fraction and depletion time depend on the inflow and ouflow rates. 
\vspace{0.2cm}

\end{abstract}


%%%%%%%%%%%%%%%%%%%%%%%%%%%%%%%%%%%%%%%%%%%%%%%%%%

%%%%%%%%%%%%%%%%% BODY OF PAPER %%%%%%%%%%%%%%%%%%

\section{Introduction}
\label{section:introduction}

%1.1 Motivations: the baryon cycle, the regulation of star formation, the main sequence, the gas-regulated quasi-equilibrium "bathtub" model, testing the relation between inflows, outflows, star formation, and gas reservoirs

One of the main open questions in galaxy evolution is related to the processes responsible for the star formation regulation in galaxies. Observations have shown that, since at least $z\sim4$, star-forming galaxies (SFG) follow a relatively tight and almost linear relation between their stellar mass ($M_\star$) and star formation rate (SFR), the so-called star formation ``main-sequence'' \citep[MS;][]{Noeske2007,Elbaz2007,Elbaz2011, Daddi2007, Schiminovich2007, Rodighiero2010, Sargent2012, Whitaker2012, Whitaker2014, Speagle2014, Renzini2015, Boogaard2018}. This redshift-dependent and robust relation promotes an overall smooth and continuous mode for star formation in galaxies, which could be sustained by a continuous supply of gas from the cosmic web and minor mergers \citep{Birnboim2003, Keres2005, Keres2009, Dekel2006, Ocvirk2008, Dekel2009, Genel2010} that provides large reservoirs of molecular gas to fuel star formation \citep{Erb2006,Daddi2010, Tacconi2010, Tacconi2013, Tacconi2018, Sargent2014, Freundlich2019,Genzel2015}. The rotating disc morphology of most galaxies that constitute this sequence \citep{ForsterSchreiber2006,Genzel2006, Genzel2008, Stark2008, Daddi2010, Wuyts2011a} and the long star formation cycles inferred from the number of SFGs observed at $z=1-2$ \citep{Daddi2005, Daddi2007, Caputi2006} further argue in favour of such a smooth mode of star formation. Typical SFGs are thought to evolve along the MS in a slowly evolving gas-regulated quasi equilibrium between inflows, outflows, and star formation \citep{Bouche2010,Dave2011a, Dave2012, Feldmann2013, Lilly2013, Dekel2013, Peng2014, Dekel2014}, until their star formation is quenched when they enter denser environment or reach a typical stellar mass of $\sim 10^{11}~M_\odot$ \citep{Conroy2009,Peng2010}. SFGs then rapidly cease their star formation and drop below the MS to populate the ``red sequence''. Before this final quenching, SFGs may oscillate within the scatter of the MS through episodes of gas compaction, enhanced star formation above the MS line, gas depletion due to star formation, limited quenching below the MS line, and replenishment by external accretion as suggested by cosmological simulations  \citep{Dekel2014,Zolotov2015,Tacchella2016a,Tacchella2016b,Dekel2019,Dekel2020a,Dekel2020b}. Both the quasi equilibrium (or ``bathtub'') model and these oscillations around the MS line imply tight relations between the gas content of SFGs, the amount of gas flowing in and out, and the star formation activity. Under the quasi equilibrium model, mass conservation notably implies a relation between gas fraction and stellar mass, and one can in principle put indirect constraints on outflow rates from molecular gas observations as in \cite{Seko2016}.

%1.2 The MEGAFLOW survey: winds and accretion, presentation of the program, the sample, the results presented so far (Schroetter+16,+19, Zabl+19a,b,...)

Observationally testing such relations is challenging, notably as signatures of gas flows are difficult to observe directly and as their location is highly uncertain when probed via standard galaxy emission or absorption lines. For example, outflow rates measured traditionally using blue-shifted absorption lines have order-of-magnitude uncertainties because the location of the absorbing gas is unknown (it can be located at 0.1, 1, or even 10 kpc away from the host galaxy). Absorption along serendipitous lines of sight of background quasars enables to significantly reduce these uncertainties by probing the circumgalactic medium (CGM) of SFGs around a given impact parameter \citep{Bouche2012,Bouche2013, Bouche2016,Kacprzak2014,Muzahid2015, Schroetter2015}.
Despite the scarcity of galaxy-quasar pairs, the unprecedented field-of-view and sensitivity of the Multi Unit Spectroscopic Explorer \citep[MUSE;][]{Bacon2006,Bacon2010} instrument together with the high-resolution Ultraviolet and Visual Echelle Spectrograph \citep[UVES;][]{Dekker2000} on the Very Large Telescope (VLT) have enabled to isolate a sample of about 100 SFGs at $z=0.5-1.4$ with a background quasar within 100~kpc as part of the MusE GAs FLOw and Wind (MEGAFLOW) survey \citep{Schroetter2016, Schroetter2019, Zabl2019, Zabl2020}. These galaxies were selected through their \ion{Mg}{II}  absorption on the lines-of-sight of 22 quasars from the \cite{Zhu2013} catalog and constitute the largest sample of \ion{Mg}{II} absorber-galaxy pairs with spectroscopic and kinematic information to date. The azimuthal angle between the quasar location and the galaxy major axis follows a clear bimodality, indicating that strong \ion{Mg}{II} absorption lies predominantly in outflow cones and extended disk-like structures \citep[cf. also][]{Bordoloi2011, Bouche2012, Kacprzak2012}. This is further corroborated by cosmological simulations, which predict more substantial inflow (outflow) detections along the galaxy major (minor) axis \citep[e.g.,][]{Peroux2020}, and by the observed correlation between metallicity and azimuthal angle, gas located along the galaxy minor axis being on average more enriched than that located along the major axis \citep{Wendt2020}. 
Pairs with azimutal angle compatible with outflows were studied in \cite{Schroetter2016}, \cite{Schroetter2019} and \cite{Zabl2020} while pairs compatible with accretion around the disc plane were studied in \cite{Zabl2019}.
There has been no attempt so far to unveil the molecular gas content, i.e., the fuel reservoir for future star formation, in the SFGs caught in the act with outflowing or accreting gas by the MEGAFLOW survey.

%1.3 Goals of the NOEMA pilot program: characterise the molecular gas content in a small but well-define sample from the MEGAFLOW survey, outflows and accretion cases, comparison to other molecular gas observational programs (PHIBSS) and scaling relations (Tacconi+18), outline of the present paper

In this paper, we present the NOEMA MEGAFLOW pilot program, a first attempt at probing molecular gas reservoirs in galaxies with gas flows drawn from the MEGAFLOW survey with IRAM's NOrthern Extended Millimeter Array \citep[NOEMA;][]{Schuster2014}. The primary objectives of this pilot program is to characterize the molecular gas content in a small but well-defined sample of galaxies with clear detections of either infalling or ouflowing gas, for which we have estimates of the gas accretion or ouflow rates, and to compare the resulting molecular gas fraction and depletion time with existing molecular gas measurements and scaling relations such as those of the PHIBSS and PHIBSS2 IRAM programs \citep{Tacconi2010, Tacconi2013, Tacconi2018, Genzel2013, Genzel2015, Freundlich2013, Freundlich2019}. These programs provide an ideal comparison sample, as their targets were selected to uniformly sample the MS and its scatter in the $M_\star$-SFR plane above given $M_\star$ and SFR thresholds, and include about 120 molecular gas measurements in the MEGAFLOW redshift range ($z= 0.5-1.4$). In particular, since a major outflow can lead to a gas depletion situation, comparing the gas fraction and depletion time of the MEGAFLOW outflow sample with those of the reference sample can validate (or not) the gas cycling event: if the gas properties of the outflow sample follow the \cite{Tacconi2018} scaling relations, it would mean that outflow events do not have a strong impact on the galaxy gas content, contrarily to the prediction from simulations; if the gas properties have smaller gas content or depend on the location on the MS, it would validate the models. 
From an initial proposed sample of 11 galaxies, the NOEMA MEGAFLOW program eventually includes only 6 targets with no clear individual detections but whose stack can be compared to the \cite{Tacconi2018} molecular gas scaling relations. This pilot program enables to lay the ground for more ambitious programs that would test the quasi equilibrium and compaction models with simultaneous individual measurements of gas flows and gas content at intermediate redshifts. 
Section~\ref{section:methods} presents the sample, the observations, and the methodology of the program; Section~\ref{section:results} the resulting upper limits, the stacking analysis, and the comparison with the scaling relations; Section~\ref{section:conclusion} concludes and discusses the results. We further present in Appendix \ref{appendix:quasar} the additional quasar continuum detections we obtained. 
Throughout this paper, we asssume a flat $\Lambda$CDM universe with $\Omega_{\rm m }=0.3$, $\Omega_{\rm \Lambda}=0.7$, and $H_0=70~\rm km~s^{-1} ~Mpc^{-1}$.

\section{Sample selection and observations}
\label{section:methods}

\subsection{The MEGAFLOW survey}

%2.1 Sample selection: drawn from the MEGAFLOW sample, characteristics of the galaxies, ancillary data (Mstar, SFR, etc), position relative to the main sequence, wind and accretion cases, initial sample vs. actually observed sample (5+1 galaxies out of 11)

The MEGAFLOW strategy consists in selecting quasar lines-of-sight with at least three \ion{Mg}{II} $\mathrm{\lambda\lambda}2797,2803$ absorbers from the \cite{Zhu2013} catalog based on the Sloan Digital Sky Survey \citep[SDSS;][]{Ross2012,Alam2015}. These \ion{Mg}{II} absorbers are further selected to have redshifts in the range $0.4-1.45$, such that the [\ion{O}{II}] $\lambda\lambda 3727,3729$ galaxy emission lines fall within the MUSE wavelength range ($4800-9300~\AA$), and rest-frame equivalent widths larger than $>0.5-0.8~\AA$ in order to have impact parameters $b\lesssim 100~\rm kpc$ given the anticorrelation between the \ion{Mg}{II} equivalent width and the impact parameter \citep{Lanzetta1990, Steidel1995, Chen2010, Bordoloi2011, Kacprzak2011b, Nielsen2013, Werk2013}. The resulting sample includes 79 strong \ion{Mg}{II} absorbers with $0.51<z<1.45$ and 38 isolated galaxy-absorber associations within 22 quasar fields.

\begin{figure*}
	\includegraphics[height=0.38\linewidth,trim={1.2cm 0cm .3cm .3cm},clip]{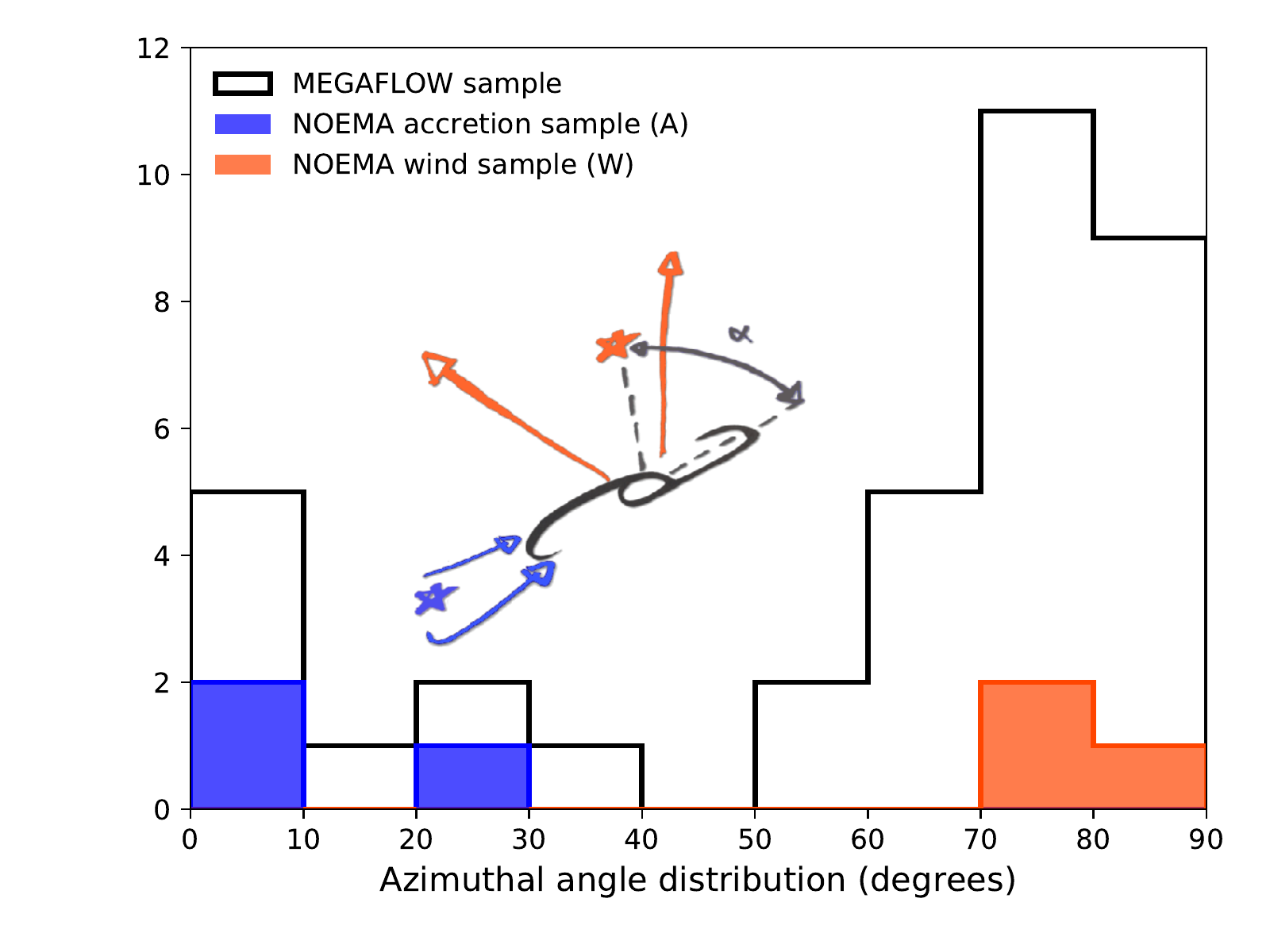}
	\hfill
	\includegraphics[height=0.38\linewidth,trim={0.1cm 0cm .3cm .3cm},clip]{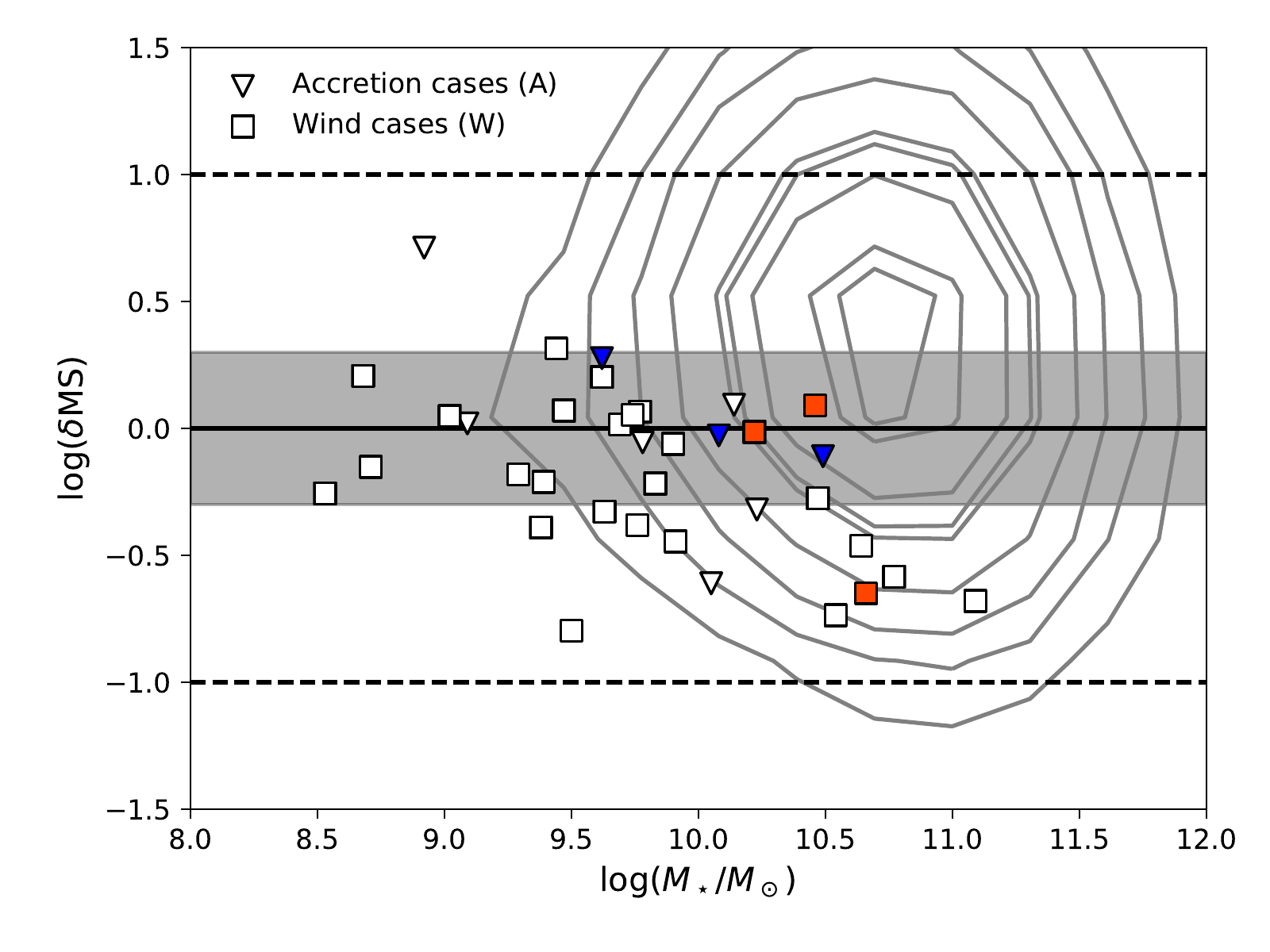}
	\vspace{-0.2cm}
	\caption{\textit{Left:} Distribution of the azimuthal angle between the galaxy major axis and the quasar line-of-sight for the entire MEGAFLOW survey (plain black line) and for the sample observed with NOEMA (blue and red filled histograms). Accretion cases are represented in blue, wind cases in red. The blue (red) star in the overlaid drawing represents the position of the quasar in an accretion (wind) configuration. \textit{Right:} Position of the MEGAFLOW galaxies with respect to the MS, where the offset from the MS ($\delta {\rm MS = SFR/SFR(MS};z,M_\star)$) is plotted as a function of the stellar mass ($M_\star$) and the reference ${\rm SFR(MS};z,M_\star)$ is that of \protect\cite{Speagle2014}. Empty triangles and squares are respectively the MEGAFLOW accretion and wind galaxies, those highlighted in blue and red correspond to the observed ones. The gray background contours indicate the \protect\cite{Tacconi2018} sample, the plain horizontal line the MS ridge ($\delta MS=1$), the shaded area its 0.3 dex scatter, and the dashed line $\pm 1$ dex from it.}
	\label{fig:sample}
\end{figure*}

As can be seen in the left panel of Fig.~\ref{fig:sample}, the distribution of the azimuthal angle ($\alpha$) between the apparent quasar location and the galaxy major axis is bimodal, indicating that strong \ion{Mg}{II} absorption is preferentially found either along the minor axis or along the major axis of the galaxies \citep{Bouche2012,Kacprzak2012, Zabl2019, Schroetter2019}. \ion{Mg}{II} absorptions along the minor axis ($|\alpha|\geq 55^\circ$) are identified with outflows \citep{Schroetter2019}, since outflows are expelling baryons from the galaxy perpendicularly to the star-forming disk in the direction of least resistance; \ion{Mg}{II} absorptions along the major axis ($|\alpha|<40^\circ$) are identified with accretion around the disc plane \citep{Zabl2019}. 

Redshifts and [\ion{O}{II}] fluxes are measured using the 3D morpho-kinematics fitting tool \texttt{Galpak3D} \citep{Bouche2015} on the MUSE data; 
stellar masses are estimated as in \cite{Schroetter2019} from the tight correlation between the stellar mass and the dynamical estimator \mbox{$S_{05}=\sqrt{0.5 \times V_{\rm max}^2 +\sigma^2}$}, which combines the rotational velocity $V_{\rm max}$ and the galaxy velocity dispersion $\sigma$ derived by \texttt{Galpak3D}, using the relation from \cite{Alcorn2018}; 
SFRs are determined from the MUSE [\ion{O}{II}] fluxes using the \cite{Kewley2004} relation corrected for extinction using the \cite{Calzetti2000} law with the strength of the extinction estimated from the $M_\star-E(B-V)$ relation of \cite{Garn2010} and assuming a \cite{Chabrier2003} initial mass function (IMF). More details can be found in \cite{Zabl2019}. Typical uncertainties on the stellar mass and the SFR are about 0.2 dex \citep[e.g.][]{Wuyts2011b, Whitaker2014,Roediger2015}.
Mass accretion and outflowing rates as well as mass loading factors can be determined from the \ion{Mg}{II} equivalent widths, the impact parameters, and the \ion{Mg}{II} kinematics \citep[cf.][]{Bouche2012, Bouche2013, Schroetter2015, Schroetter2016, Schroetter2019, Zabl2019}. 

The right panel of Fig.~\ref{fig:sample} shows the location of the MEGAFLOW galaxies with respect to the MS in terms of their SFR offset from the MS, 
\be
\delta {\rm MS} = {\rm SFR/SFR(MS};z,M_\star)
\ee
where ${\rm SFR(MS};z,M_\star)$ is the analytical prescription for the center of the MS as a function of redshift and stellar mass proposed in the compilation by \cite{Speagle2014}. 
All MEGAFLOW galaxies lie within the scatter of the MS, although galaxies above $\log(M_{\rm star}/M_\odot)\sim 10$, which are dominated by wind cases, tend to be slightly below the \citet{Speagle2014} MS line.
This may suggest that outflows above this mass may induce a star formation decrease. 
This potential relation between
$\delta$MS and gas flows  will be investigated further with the full survey (Bouché et al., in prep.). 

While the average $\log(M_{\rm star}/M_\odot)$ and $\delta {\rm MS}$ of the \cite{Tacconi2018} sample are respectively 10.7 and 0.39, those of the MEGAFLOW sample are 9.8 and $-0.16$ and those of the sample observed with NOEMA 10.3 and $-0.07$: the MEGAFLOW sample and the NOEMA subsample have lower stellar masses and offsets from the MS than the \cite{Tacconi2018} sample on average.

\subsection{The pilot program sample}

For the NOEMA MEGAFLOW pilot program presented here, we selected a subset of SFGs from the MEGAFLOW sample with measured accretion and outflow events, mass flux estimates, stellar masses, and SFRs. 
The sources were selected to optimize the NOEMA CO(4-3) or CO(3-2) observing times given their SFR, stellar masses, and redshifts, assuming the \cite{Tacconi2018} scaling relations to assess their expected line fluxes. As such, the sample is notably biased towards higher SFRs.
As shown in Fig.~\ref{fig:sample}, the current sub-sample comprises 6 SFGs equally divided between accretion and wind cases.
Table~\ref{table:sample} summarizes some of the properties of the targets, including their R.A., DEC. coordinates, redshifts, stellar masses, SFRs, [\ion{O}{II}] line widths, and \ion{Mg}{II} rest-frame equivalent width in quasar sightline.
Each full galaxy ID includes the quasar identifier, the absorber redshift, the impact parameter in arcsec, and the position angle between quasar and galaxy in degrees. In the following, we use the absorber redshift as short ID. 
The positions and redshifts of the quasars are indicated in Table~\ref{table:quasars}.

\begin{table*}
	\caption{Parameters of the six galaxies observed as part of the NOEMA MEGAFLOW pilot program. 
		(1) Galaxy ID, with the short ID highlighted in bold. 
		(2) Type: A for an accretion case ($|\alpha|< 40^\circ$), W for a wind case ($|\alpha|\geq 55^\circ$). 
		(3)(4) Spatial coordinates. 
		(5)(6)(7)(8)(9) Redshift, stellar mass, SFR, [\ion{O}{II}] FWHM, and \ion{Mg}{II} rest-frame equivalent width in quasar sightline \protect\citep[cf.][]{Zabl2019, Schroetter2019}. 
		}
	\label{table:sample}
	\centering
	\resizebox{\textwidth}{!}{
		\begin{tabular}{lcccccccccrr}
			\hline\hline
			\noalign{\vskip 1mm}  ID  & Type & R.A. & DEC. & $z$ & $\rm \log(M_{star}/M_\odot)$ & $\rm SFR$ $\rm [M_\odot yr^{-1}]$ & FWHM [km/s]  & $\rewmgii$ $\rm [\AA$] \\
			(1) & (2)  & (3) & (4) & (5)  & (6) & (7) & (8)& (9)\\
			\hline
			\noalign{\vskip 1mm}
			%23
			gal\_J0800p1849\_\textbf{0608}\_10\_108 &A&08:00:05.200 &+18:49:32.550&   0.6082 & 9.62 & 4.99 & 180  & 0.8 \\
			%52 
			gal\_J1236p0725\_\textbf{0912}\_2\_246 &A&12:36:24.250 &+07:25:50.580&   0.9128 & 10.49 & 12.96 & 375  & 2.2 \\
			%39 
			gal\_J1039p0714\_\textbf{0949}\_6\_324 &A&10:39:36.420 &+07:14:32.370&   0.9494 & 10.08 & 8.78 & 235 & 1.2 \\
			\hline
			\noalign{\vskip 1mm}
			%32 
			gal\_J0838p0257\_\textbf{1099}\_8\_160 &W&08:38:52.200 &+02:56:56.600&   1.1001 & 10.22 & 13.83 & 170 & 0.1 \\
			%33 
			gal\_J0937p0656\_\textbf{0702}\_6\_209 &W&09:37:49.400 &+06:56:51.640& 0.7021 & 10.46 & 13.42 & 270  & 1.8 \\
			%188
			gal\_J0015m0751\_\textbf{0731}\_5\_3&W& 00:15:35.190 &-07:50:57.890& 0.7313 & 10.66 & 3.49 & 217& 2.1 \\
			\hline
			%\hline
		\end{tabular}
	}
\end{table*}

\subsection{Observations and data reduction}
\label{section:observations+reduction}

%2.2 NOEMA observations and data reduction: when the observations were carried out, which programs, which bands, which spectral/array configurations, observational times, one galaxy only partially observed, weather conditions, data reduction details

To probe the molecular gas content of the sample galaxies, we observe either the CO(3-2) or the CO(4-3) carbon monoxide emission lines ($\nu_{\rm rest}=345.8$ and $461.0~\rm GHz$, respectively), which fall in NOEMA's 1.3mm band given the redshift range of the targets and are both located near the peak of the CO line SED for SFGs \citep[e.g.,][]{Daddi2015}. 
The six galaxies were observed at the NOEMA interferometer at the Plateau de Bure in France between March and June 2019, as part of the observational program W18CS (P.I.: N. Bouch\'e, T. Contini). The CO(3-2) and CO(4-3) emission lines from the targets were observed at frequencies between 203 and 266 GHz, falling in the 3mm band, in the extended D configuration with 10 antennas until beginning April 2019 and 9 afterwards. The beam size, the eventual on-source observing time for each target as well as the resulting RMS noise are indicated in Table~\ref{table:CO}. %The total on-source time is 13.6 hours. 
The weather conditions varied from good to mediocre, with system temperatures ranging between 100 and 500 K depending on atmospheric conditions, wind speeds between 1 and 13m/s, and water vapor between 0.5 and 5mm. The absolute flux scale was derived from secondary flux calibrators (LkH$\alpha$101, 0735+178, 0923+392, MWC349, 0829+046, and 3C454.3), whose fluxes are regularly measured using Jupiter satellites or planets. The data were calibrated using the \texttt{clic} software of the IRAM \texttt{gildas} package, and further analysed and mapped using the \texttt{gildas} \texttt{mapping} and \texttt{class} sofwares. 

\begin{table}
	\caption{CO observations. 
		(1) Short ID, i.e., the identifier of the absorber redshift in the full galaxy ID quoted in Table~\ref{table:sample}. 
		(2) Targetted CO transition. 
		(3) Beam size. 
		(4) Total on-source time.
		(5) RMS noise over the FWTM. 
		}
	\label{table:CO}
	\centering
	%\resizebox{\textwidth}{!}{
		\begin{tabular}{llccccccccccccrr}
			\hline\hline
			\noalign{\vskip 1mm}  ID  &  CO line & beam  & $\rm t_{obs}$ $\rm [h]$ & $\rm \sigma_{\rm FWTM}$ $\rm [Jy~km/s]$ \\
			(1) & (2)  & (3) & (4) & (5) \\
			\hline
			\noalign{\vskip 1mm}
			%23
			0608 & 3-2 & $2.55"\times2.15"$ & 4.1 & 0.11 \\
			%52 
			0912 & 4-3  & $2.05"\times1.24"$ & 0.8 & 0.17 \\
			%39 
			0949 & 4-3  & $3.09"\times1.51"$ & 3.0 & 0.19 \\
			\hline
			\noalign{\vskip 1mm}
			%32 
			1099 & 4-3 & $2.39"\times 1.60"$ & 0.9 & 0.17 \\
			%33 
			0702 & 3-2  & $2.99"\times1.76" $ & 1.4 & 0.15 \\
			%188
			0731 & 4-3 & $3.13"\times1.47"$ & 3.4 & 0.24  \\
			\hline
			%\hline
		\end{tabular}
	%}
\end{table}

\section{Results}
\label{section:results}

\subsection{CO molecular gas}
\label{section:CO}

%3.1 Molecular gas properties: spectra, upper limits, conversion to molecular gas masses, result table, possible hint of detection with low SNR for ID188

%3.4 Comparison with the Tacconi+18 sample: KS relation, molecular gas fraction, depletion time, stacked value which seems to fall on the expected value from the Tacconi+18 relation, check if possible to get stacks separately for wind and accretion cases (not sure, but we can try)

With beam sizes (Table~\ref{table:CO}) at least 2-3 times larger than the average galaxy size on the MS at the redshift and stellar mass of the sources \citep{Vanderwel2014}, our NOEMA observations are unresolved. 
For each observed galaxy, we extract the spectrum at the optical position by fitting the point spread function (PSF) in the $uv$ Fourier space using \texttt{gildas mapping go uvfit} tool. 
Fitting the unresolved sources through their $uv$ visibilities avoids the imaging and deconvolution steps, which involve assumptions to compensate for the sparse $uv$ coverage.
The CO(3-2) and CO(4-3) spectra of the six galaxies of the sample are displayed in Fig.~\ref{fig:spectra}. We use the MUSE [\ion{O}{II}] FWHM to evaluate the full width at tenth-maximum (FWTM) of the expected CO line assuming a Gaussian profile, i.e., $\rm FWTM=FWHM \times \sqrt{\log(10)/\log(2)}$, and evaluate both the signal and the noise within the FWTM -- which is expected to contain 97\% of the flux. 
Indeed, since the targetted galaxies are typical MS SFGs, obscuration is not extreme and both [\ion{O}{II}] and CO should trace similar star-forming regions and velocity fields. Observational studies have also shown that [\ion{O}{II}] and CO line widths were comparable \citep[e.g.,][Puglisi et al. 2020, Nature Astronomy in press]{Freundlich2013}.
The RMS noise indicated in Table~\ref{table:CO} were obtained using \texttt{gildas} \texttt{mapping} \texttt{go noise} tool over velocity channels of width $\sim50\rm~km/s$, rescaled to the [\ion{O}{II}] FWTM.  None of the galaxies is detected, although there may be hints of detection at low signal-to-noise ratio (SNR) for galaxies 0702 and 0731. For these two galaxies, the velocity integrated line fluxes within the FWTM are respectively $\rm 0.280 ~Jy~km/s$ ($\rm SNR=1.9$) and $\rm 0.567~Jy~km/s$ ($\rm SNR=2.4$). Given the low SNR, we prefer to treat these measurements as non-detections. We nevertheless highlight that these fluxes and SNR estimates are obtained without any free parameters, since the positions and widths of the expected lines are set. 
We define $3\sigma$ integrated line flux upper limits as $F_{\rm upper}= 3\times \sigma_{\rm FWTM}$, which correspond to the fluxes that should have been detected at $\rm SNR>3$  with a 50\% probability \citep[e.g.,][]{Masci2011}. 

From any velocity integrated CO($J\rightarrow J-1$) transition line flux $F_{{\rm CO}(J\rightarrow J-1)}$ (or upper limits $F_{\rm upper}$), one can derive the intrinsic CO luminosity 
\begin{equation}
\label{eq:luminosity}
\left(\frac{L^\prime_{{\rm CO}(J\rightarrow J-1)}}{\mathrm{K~km~s}^{-1}\mathrm{~pc}^2}\right) = \frac{3.25 \times 10^7}{1+z} 
\left(\frac{F_{{\rm CO}(J\rightarrow J-1)}}{\mathrm{Jy~km~s}^{-1}}\right)
\left(\frac{\nu_{\rm rest}}{\mathrm{GHz}}\right)^{-2} \left(\frac{D_\mathrm{L}}{\mathrm{Mpc}}\right)^2,
\end{equation}
where $\nu_{\rm rest}$ is the rest-frame frequency ($\rm 345.8 ~GHz$ for \mbox{CO(3-2)}, $\rm 461.041.8 ~GHz$ for \mbox{CO(4-3)}), and $D_{\rm L}$ the luminosity distance \citep{Solomon1997}. 
Even if the CO molecule only represents a small fraction of the total molecular gas mass and if its lower rotational lines are almost always optically thick, this quantity can be used as a quantitative tracer of the molecular gas mass, estimated as 
\begin{equation}
M_{\rm gas} = \alpha_{\rm CO} L_{{\rm CO}(J\rightarrow J-1)}^\prime / r_{\rm J1},
\end{equation}
where $\alpha_{\rm CO}$ is the CO(1-0) luminosity-to-molecular-gas-mass conversion factor and $r_{J1}=L_{{\rm CO}(J\rightarrow J-1)}^\prime/L_{CO(1-0)}^\prime$ the corresponding line ratio. 
Since CO emission in the $z=0.6-1.1$ galaxies studied in this paper is likely to originate from virialized giant molecular clouds with mean densities of the same order of magnitude as their lower-redshift counterparts and similar dust temperatures, the conversion factor should be relatively close to the Galactic conversion factor $\alpha_G = 4.36 \pm 0.9 \rm~ M_\odot/(K~km~s^{-1} pc^{2})$.
To estimate molecular gas mass upper limits from the \mbox{CO(3-2)} and \mbox{CO(4-3)} upper limits, we follow the PHIBSS2 methodology \citep{Genzel2015, Tacconi2018,Freundlich2019}. Namely, we include a metallicity dependence of the conversion factor, taken as the geometric mean of the recipes by \cite{Bolatto2013} and \cite{Genzel2012}, 
\begin{equation}
\alpha_{\rm CO} = \alpha_G \sqrt{0.67 \times \exp(0.36 \times 10^{8.67-\log{Z}}) \times 10^{-1.27\times (\log{Z}-8.67)}}
,\end{equation}
where $\log{Z} = 12+\log({\rm O/H})$ is the metallicity on the \cite{Pettini2004} scale estimated from the mass--metallicity relation 
\begin{equation}
\log{Z} = 8.74 - 0.087 \times (\log({\rm M_{\star}})-b)^2
,\end{equation}
with $b = 10.4 + 4.46 \times \log(1+z)-1.78 \times (\log(1+z))^2$ \citep[][and references therein]{Genzel2015}. This metallicity correction leads to a mean $\alpha_{\rm CO} = 4.4 \pm 0.3 \rm~ M_\odot/(K~km~s^{-1} pc^{2})$ within the sample. 
Still following the PHIBSS2 methodology, we assume $r_{31}=0.56$ and $r_{41}=0.42$ as suggested by observations in low- and high-redshift SFGs \citep[e.g.,][]{Weiss2007, Dannerbauer2009, Bothwell2013, Bolatto2015, Daddi2015}. 
From the molecular gas mass upper limits, we derive upper limits for the molecular gas-to-stellar mass ratio $\mu_{\rm gas}=M_{\rm gas}/M_\star$ and depletion time $t_{\rm depl}=M_{\rm gas}/{\rm SFR}$. 
Table~\ref{table:CO_derived} indicates the quantities derived from the CO molecular gas measurements.

\begin{table}
	\caption{Quantities derived from the CO molecular gas measurements: intrinsic CO(1-0) luminosity, molecular gas mass, gas-to-stellar mass ratio, and depletion time. The table includes both the individual upper limits and the stacked values for all galaxies, for the accretion cases (A), and for the wind cases (W). Galaxies are referred to with their short ID (cf. Table~\ref{fig:sample}).}
	\centering
	\resizebox{\linewidth}{!}{
	\begin{tabular}{lllll}
		\hline
		\hline
		\noalign{\vskip 1mm}
		ID & $L^\prime_{\rm CO(1-0)} \rm [K~\! km\!~s^{-1}pc^2]$ & $M_{\rm gas} [M_\odot]$ & $\mu_{\rm gas}$ & $t_{\rm depl} \rm [Gyr]$\\
		\noalign{\vskip 1mm}
		\hline
		\noalign{\vskip 1mm}
		0608 &    <$1.3~10^9$  &   <$7.5~10^9$      &    <$1.80$   & <$1.5$\\
		0912 &    <$3.4~10^9$  &   <$1.50~10^{10}$   &    <$0.49$   & <$1.2$\\
		0949 &    <$4.1~10^9$  &   <$2.19~10^{10}$   &    <$1.82$   & <$2.5$\\
		\hline
		\noalign{\vskip 1mm}
		1099 &    <$4.9~10^9$  &   <$2.58~10^{10}$   &    <$1.55$   & <$1.9$\\
		0702 &    <$2.3~10^9$  &   <$9.96~10^{9}$   &    <$0.35$   & <$0.7$\\
		0731&    <$3.1~10^9$  &   <$1.24~10^{10}$   &    <$0.27$   & <$3.6$\\
		\hline
		\noalign{\vskip 1mm}
		stack & $1.4~10^9$ & $6.96~10^{9}$ & $0.39$ & $0.7$\\
		A stack & <$1.6~10^9$& <$8.24~10^{9}$ & <$0.71$ & <$0.9$ \\
		W stack & $2.1~10^9$ & $9.42~10^{9}$ & $0.34$ & $0.9$ \\
		\hline
		%\hline
	\end{tabular}
	}
	\label{table:CO_derived}
\end{table}

\begin{figure}
	\centering
	\includegraphics[width=1\linewidth,trim={0.cm 0.6cm .4cm .cm},clip]{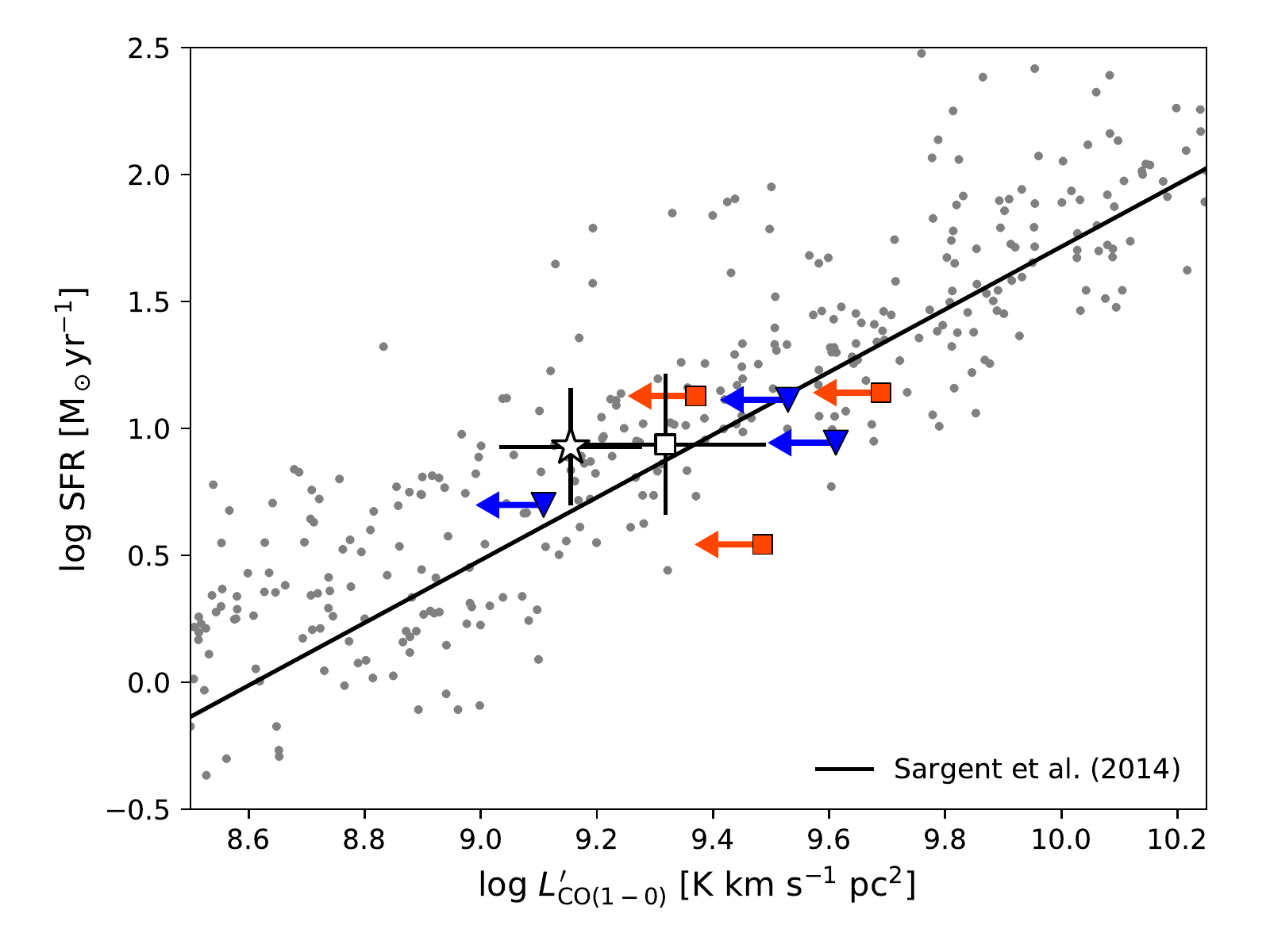}
	\vspace{-0.2cm}
	\caption{Observed relation between the SFR and the intrinsic CO(1-0) luminosity for the six galaxies of the sample (plain blue triangles and red squares) and the \protect\cite{Tacconi2018} sample of SFGs within 1 dex of the MS line (gray points). Blue triangles correspond to accretion cases, red squares to outflow cases. The empty star with error bars corresponds to the stack of all the sample galaxies; the empty square with error bars to the stack of outflow cases. 
	The solid black line corresponds to the \protect\cite{Sargent2014} fitting function. Our measurements lie within the scatter of the relation.}
	\label{fig:SFR_vs_LCO10}
\end{figure}

Fig.~\ref{fig:SFR_vs_LCO10} shows how the CO(1-0) intrinsic luminosity directly derived from the observations (assuming $r_{31}$ and $r_{41}$ as indicated above) correlates with the [\ion{O}{II}] SFR and compare with the correlation observed for SFGs within the larger \cite{Tacconi2018} sample and the fitting function proposed by \cite{Sargent2014}. The \cite{Tacconi2018} samples includes about 650 CO molecular gas measurements between redshift $z=0$ and $3.4$, from which we selected those within 1 dex of the MS line ($0.1<\delta {\rm MS}<10$), while the previous \cite{Sargent2014} compilation included 131 of these measurements. 
Our $3\sigma$ $L_{\rm CO(1-0)}$ upper limits and the stacked values with their error bars (cf. Section~\ref{section:stacks}) are compatible with the scatter of the relation.

\subsection{Comparison to existing molecular gas scaling relations}

\begin{figure*}
	\centering
	\includegraphics[width=0.49\linewidth,trim={0.cm 0.6cm .4cm .cm},clip]{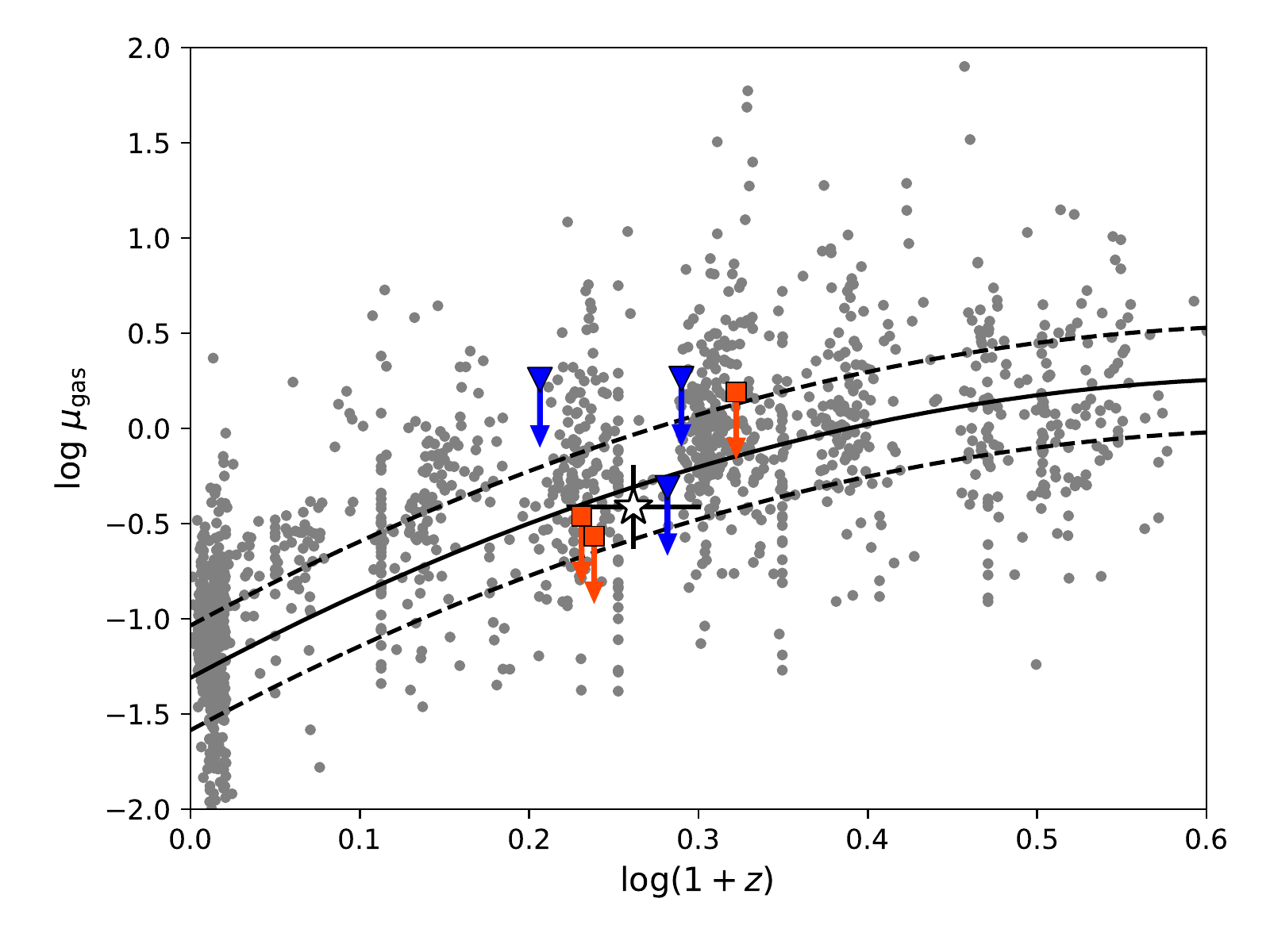}
	\hfill
	\includegraphics[width=0.49\linewidth,trim={0.cm 0.6cm .4cm .cm},clip]{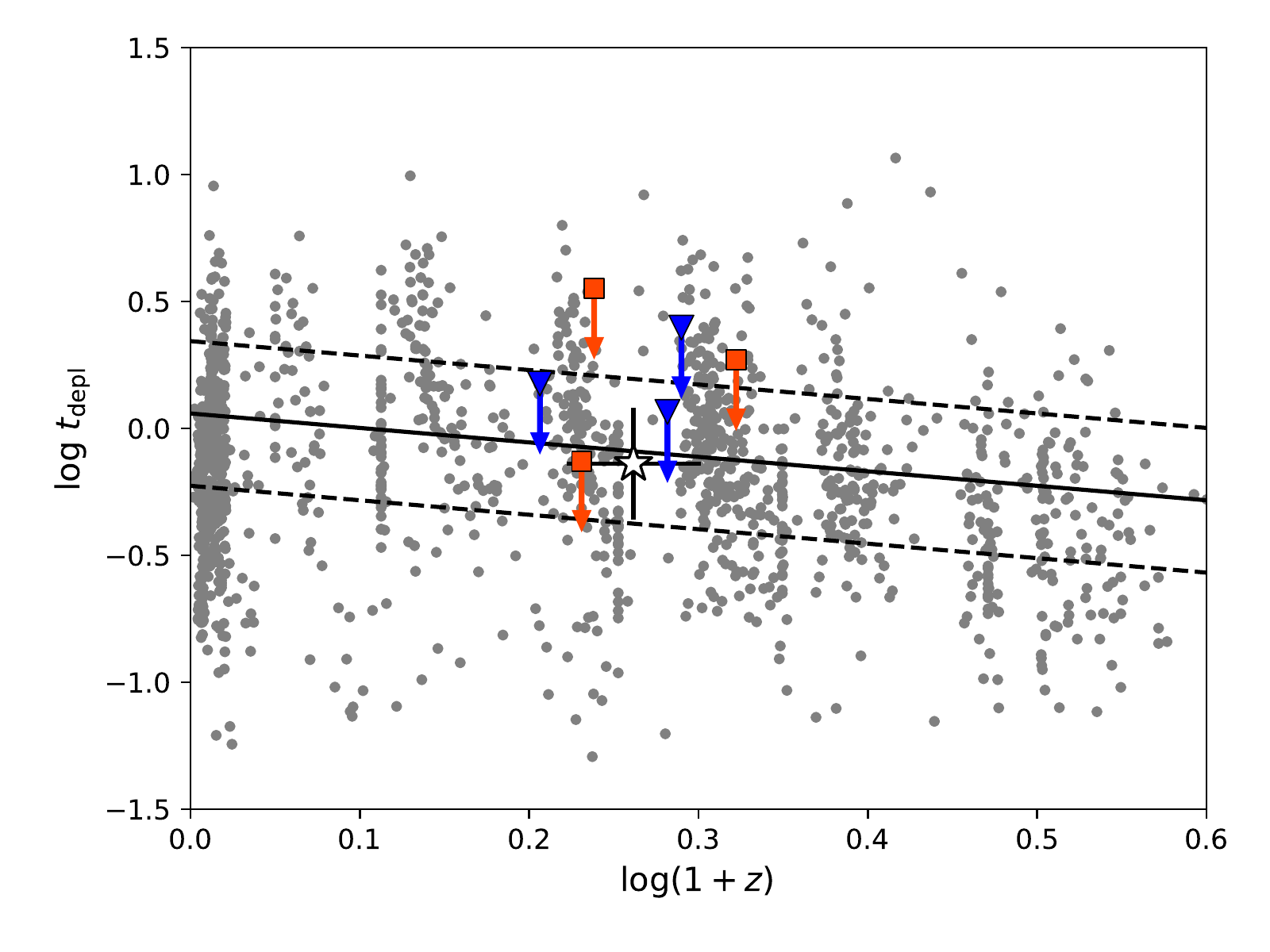}
	\vspace{-0.2cm}
	\caption{Molecular gas-to-stellar mass ratio and depletion time as a function of redshift for the six galaxies of the sample (plain blue triangles and red squares) compared to the \protect\cite{Tacconi2018} sample (gray points).  Blue triangles correspond to accretion cases, red squares to outflow cases. The empty star with error bars corresponds to the stack of all the sample galaxies. The plain black lines correspond to the \protect\cite{Tacconi2018} scaling relations for galaxies of the mean stellar mass of the sample ($\log(M_\star/M_\odot)=10.3$) on the MS ($\delta{\rm MS}=1$), the dashed lines to their scatter. Since the average mass in each redshift bin of the \protect\cite{Tacconi2018} sample may differ from $\log(M_\star/M_\odot)=10.3$, the lines are slightly offsetted compared to the gray points. With an average offset from the MS of $-0.07$ dex, the stacked detection corresponds to a molecular gas-to-stellar mass ratio slightly below the MS line.}
	\label{fig:redshift}
\end{figure*}

\begin{figure*}
	\centering
	\includegraphics[width=0.49\linewidth,trim={0.cm 0.6cm .4cm .cm},clip]{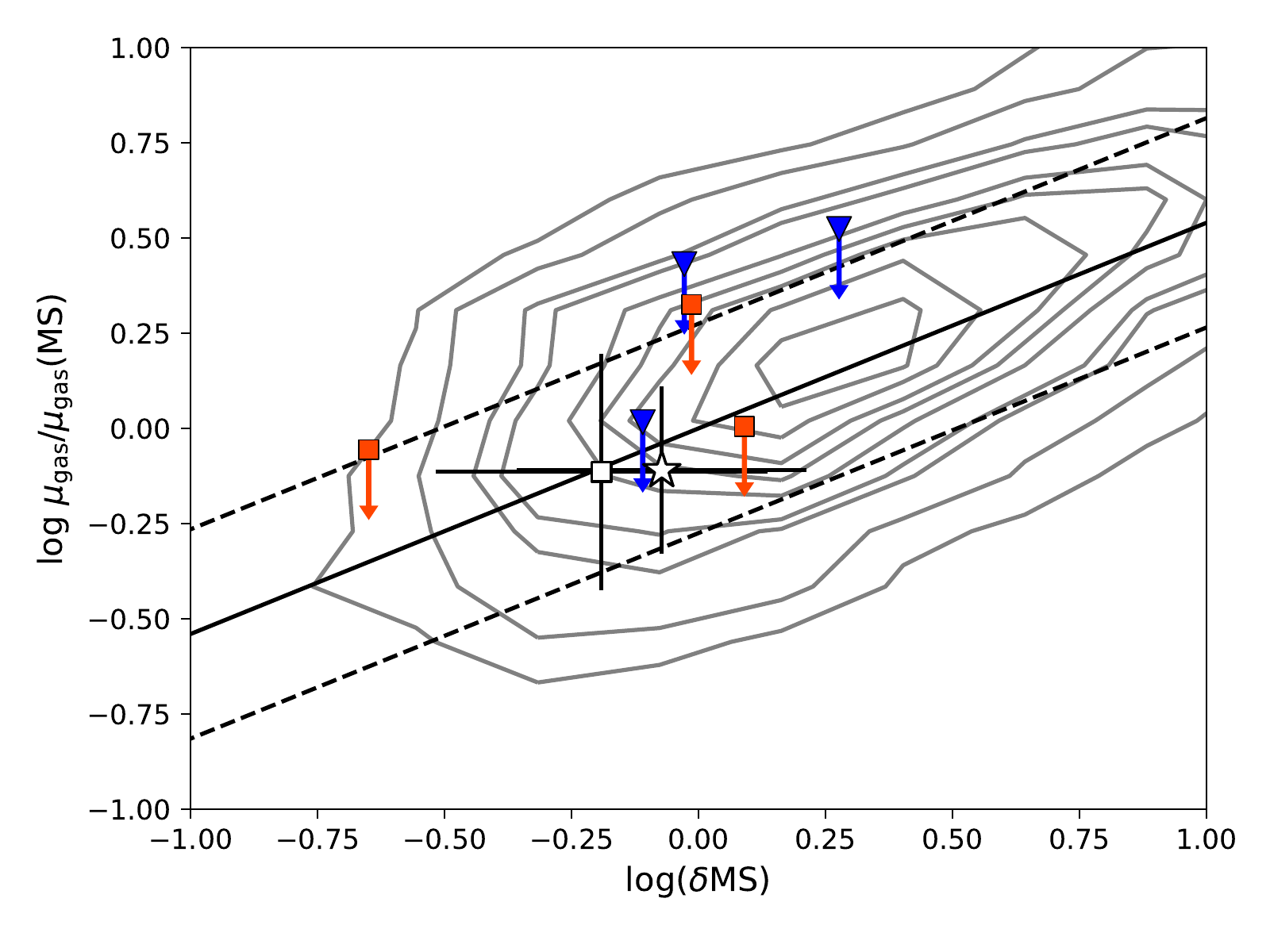}
	\hfill
	\includegraphics[width=0.49\linewidth,trim={0.cm 0.6cm .4cm .cm},clip]{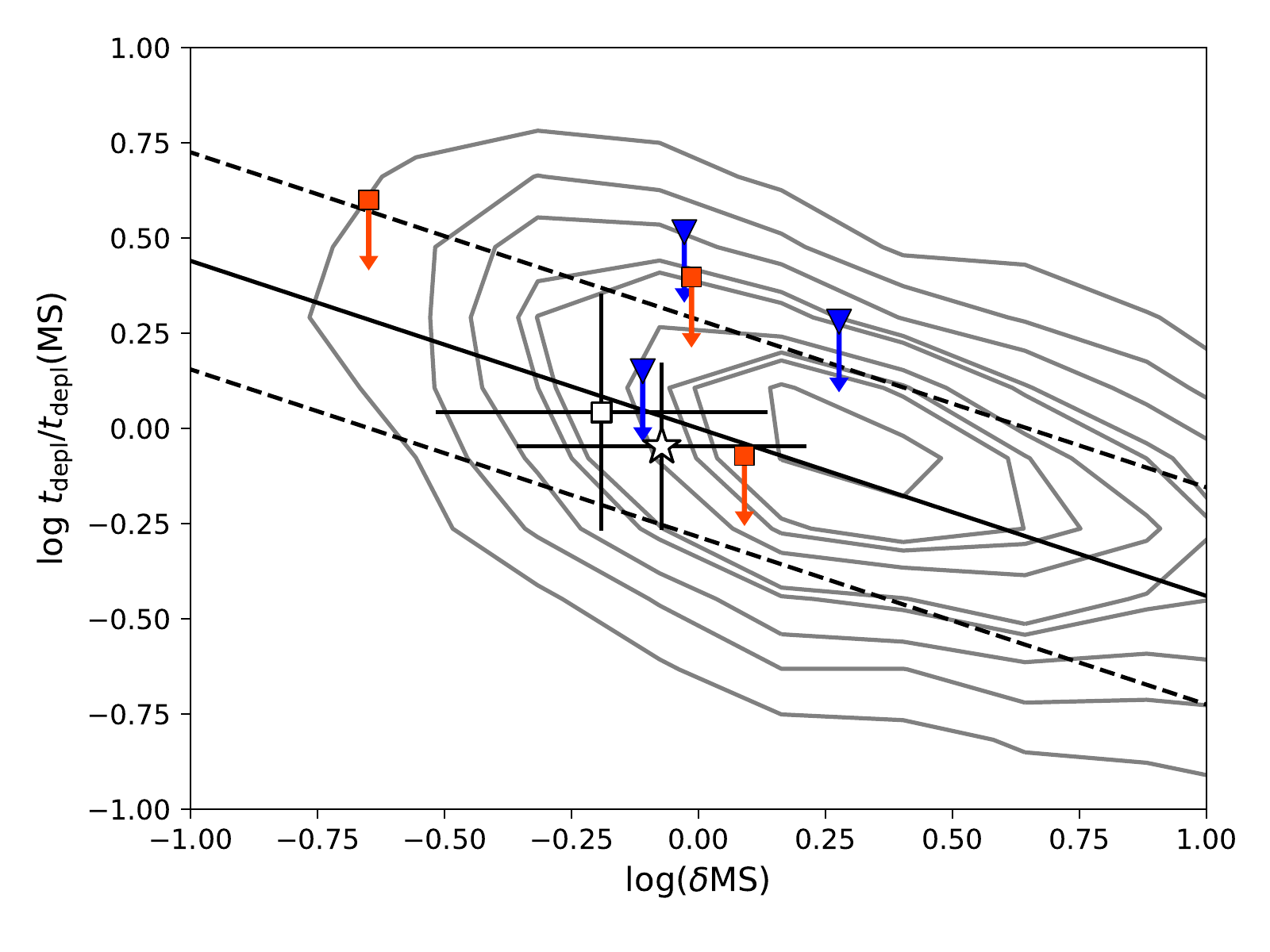}
	\vspace{-0.2cm}
	\caption{Molecular gas-to-stellar mass ratio and depletion time divided by their average values on the MS given the redshifts and stellar masses of the sources according to the \protect\cite{Tacconi2018} scaling relations as a function of the sSFR offset from the MS, $\rm \delta MS=  sSFR/sSFR(MS)$. Blue triangles correspond to accretion cases, red squares to outflow cases. The empty star with error bars corresponds to the stack of all the sample galaxies; the empty square with error bars to the stack of outflow cases. The gray background contours, the black lines, and the dashed lines respectively show the \protect\cite{Tacconi2018} sample, scaling relations, and scatter.}
	\label{fig:dMS}
\end{figure*}

Figs.~\ref{fig:redshift} and \ref{fig:dMS} compare the molecular gas-to-stellar mass ratio ($\mu_{\rm gas}=M_{\rm gas}/M_\star$) and depletion time ($t_{\rm depl}=M_{\rm gas}/{\rm SFR}$) upper limits for the six galaxies of the sample to the \cite{Tacconi2018} sample and scaling relations.
These scaling relations express $\mu_{\rm gas}$ and $t_{\rm depl}$ as a function of redshift $z$, stellar mass $M_\star$, and the specific star formation rate (${\rm sSFR}={\rm SFR}/M_\star $) offset from the MS, $\delta {\rm MS} = {\rm sSFR/sSFR}({\rm MS};z,M_\star)$, where ${\rm sSFR}({\rm MS};z,M_\star)$ is the mean sSFR on the MS given $z$ and $M_\star$. We use here the  \cite{Speagle2014} parametrisation of  ${\rm sSFR}({\rm MS};z,M_\star)$. These relations were established on a very large sample of about 1400 CO and dust molecular gas measurements in the range $z=0-4.5$, owing notably to the COLDGASS \citep{Saintonge2011,Saintonge2011b}, PHIBSS \citep{Tacconi2010, Tacconi2013}, and PHIBSS2 \citep{Freundlich2019} programs. They notably quantify how the molecular gas fraction decreases steeply with time while the depletion time increases slightly such that the cosmic evolution of the SFR is mainly driven by the available molecular gas reservoirs. They also show how the gas fraction increases above the MS and decreases below, while the depletion time follows the opposite trend. 

Fig.~\ref{fig:redshift} shows the redshift evolution of $\mu_{\rm gas}$ and $t_{\rm depl}$ from the \cite{Tacconi2018} sample\footnote{We highlight that the scaling relation lines shown in Fig.~\ref{fig:redshift} were scaled to the average stellar mass of the sample and hence are slightly offsetted from the \cite{Tacconi2018} data.} together with the upper limits we obtain in the redshift range $z=0.6-1.1$.
Fig.~\ref{fig:dMS} further shows $\mu_{\rm gas}/\mu_{\rm gas}({\rm MS})$ and $t_{\rm depl}/t_{\rm depl}({\rm MS})$ as a function of $\delta{\rm MS}$, where $\mu_{\rm gas}({\rm MS})$ and $t_{\rm depl}({\rm MS})$ are the molecular gas-to-stellar mass ratio and depletion time on the MS line ($\delta{\rm MS}=1$) according to the scaling relations given the redshift and stellar mass of the sources. The expected dependence as a function of $\delta{\rm MS}$ and its scatter is shown as the plain and dashed black lines. The upper limits are compatible with the scaling relations, and there is no clear difference between the accretion and outflow subsamples (respectively in blue and red on the figure). More precisely, if we assume a Gaussian scatter in the \cite{Tacconi2018} scaling relations, we can assess the probability for six data points following the corresponding distribution to be simultaneously below the upper limits we obtain, and we find a probability of about 60\%.

\subsection{Stacked CO analysis}
\label{section:stacks}

%3.2 Stacked spectrum

\begin{figure*}
	\centering
	\includegraphics[width=0.49\linewidth,trim={0.1cm 0cm 1.3cm 1.2cm},clip]{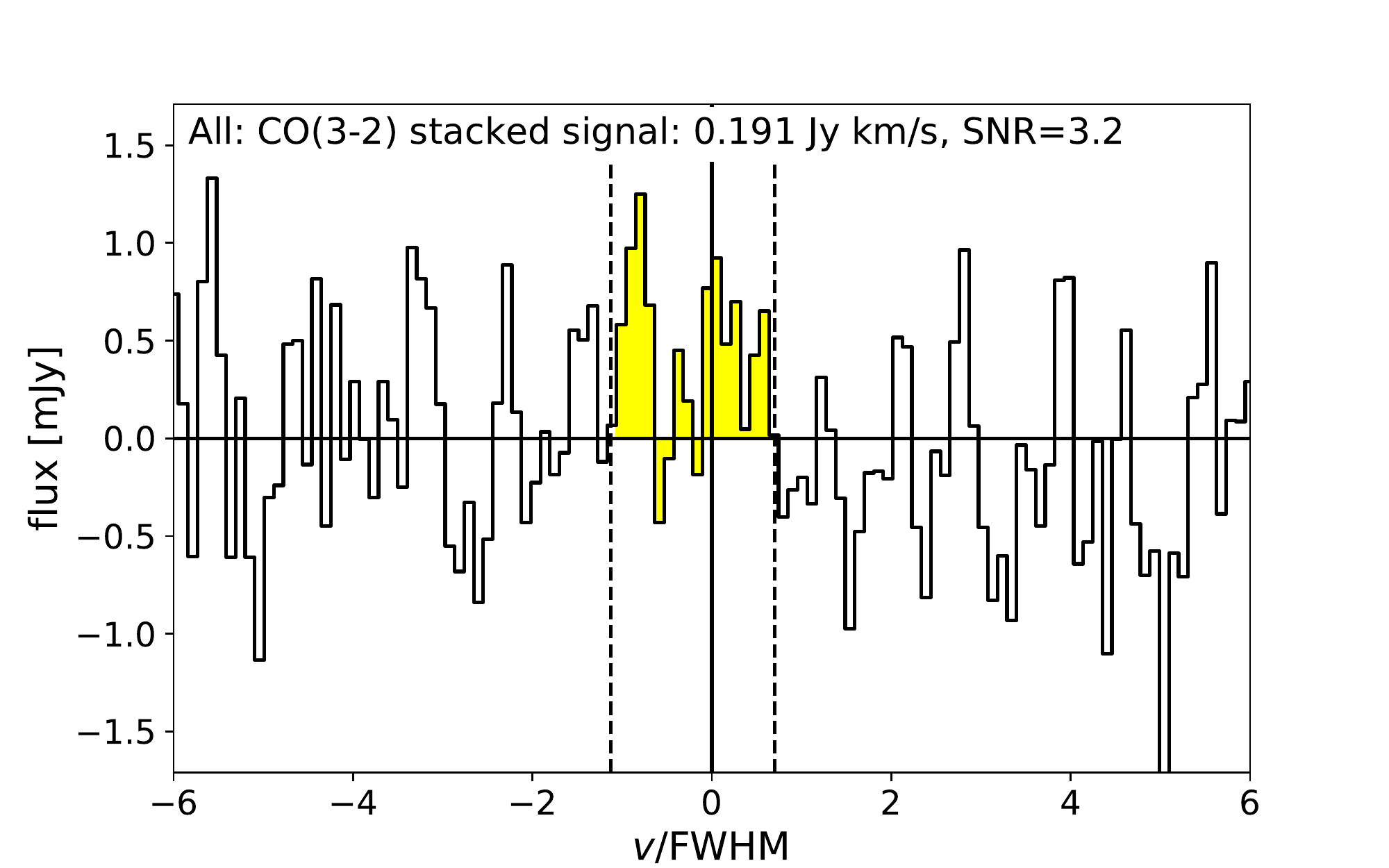}
	\includegraphics[width=0.49\linewidth,trim={0.1cm 0.cm 1.3cm 1.2cm},clip]{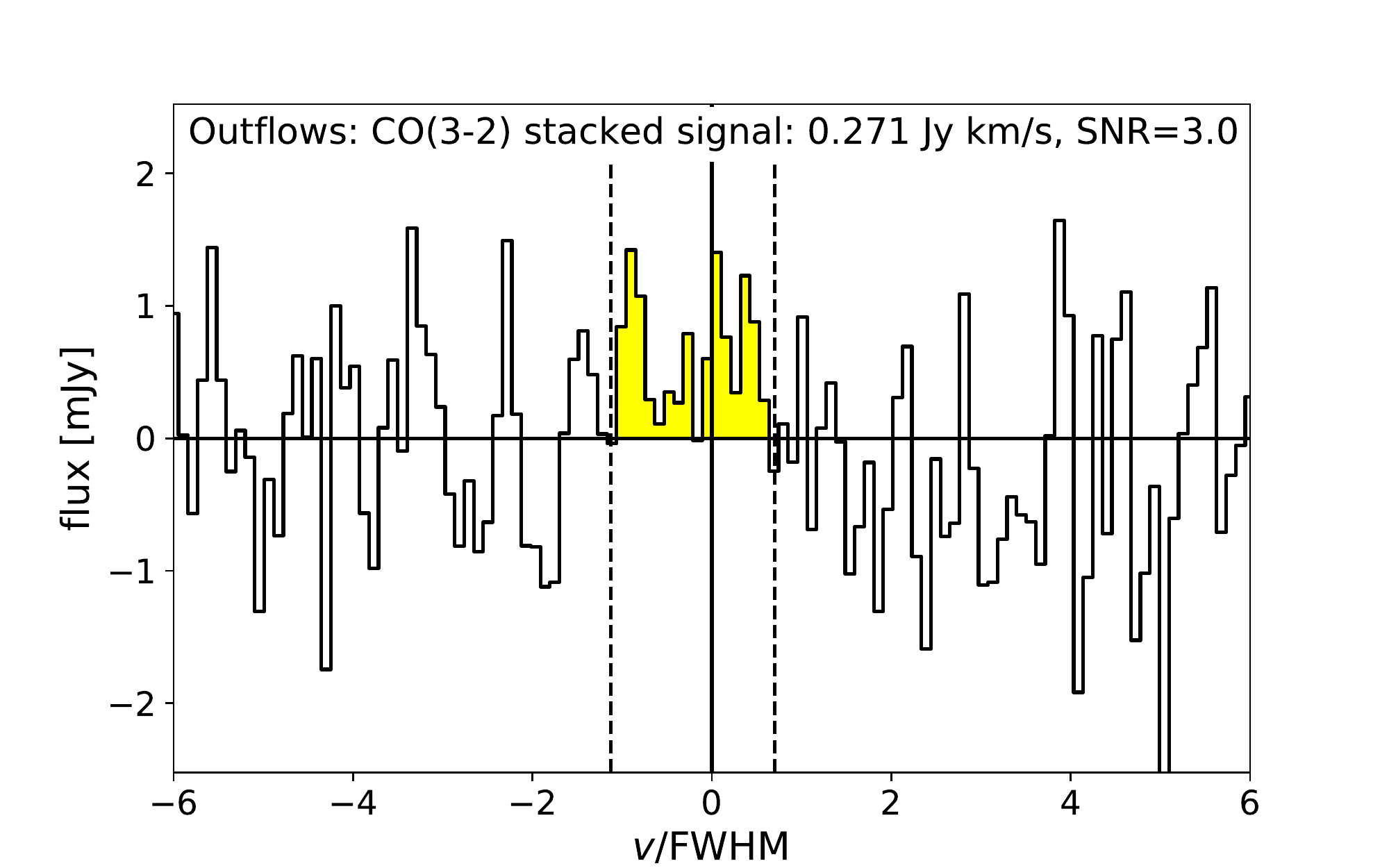}
	\vspace{-0.2cm}
		\caption{Stacked CO spectra over all galaxies (\textit{left}) and over the three outflow cases (\textit{right}). The CO(4-3) spectra were rescaled to CO(3-2) assuming a factor $r_{31}/r_{41}=1.3$ between the two line fluxes. The velocity axes were scaled to the FWHM before being stacked and we searched for a signal within the FWTM, i.e., within $\rm FWHM\times \sqrt{\log(10)/\log(2)}$, allowing a minor offset. All spectra were given the same weight. The vertical lines indicate the FWTM, the horizontal ones the flux standard deviation used to determine the SNR. For all galaxies of the sample, we obtain a signal of $0.191~\rm Jy~km/s$ (SNR=3.2), for the outflow cases of  $0.271~\rm Jy~km/s$ (SNR=3.0). We did not detect any signal for the stack of the accretion cases, which corresponds to a $0.217~\rm Jy~km/s$ 3$\sigma$ upper limit. 
		}
	\label{fig:stacks}
\end{figure*}

We stack the CO spectra at the optical positions of the six observed galaxies, rescaling the CO(4-3) fluxes to CO(3-2) fluxes assuming a factor $1.3$ between the two line fluxes and normalising the velocities by the MUSE [\ion{O}{II}] FWHM. We search for a signal within the FWTM, i.e., within $\rm FWHM\times \sqrt{\log(10)/\log(2)}$, where 97\% of the line flux should be. The resulting stacked spectra for all the galaxies of the sample and for the outflow cases are shown in Fig.~\ref{fig:stacks}. In these two cases, we detect velocity-integrated line fluxes respectively of $0.191~\rm Jy~km/s$ and $0.271~\rm Jy~km/s$ with $\rm SNR=3.2$ and $3.0$, noise being evaluated from the standard deviation of the stacked flux over the FWTM.
For the accretion cases, the stacked spectrum does not display any detection (the SNR would be $\sim$1.5) but enables to place a 3$\sigma$ upper limit of $0.217~\rm Jy~km/s$. 
The 3$\sigma$ detections are obtained almost without any free parameter since both the central velocities and the line widths are contrained by the MUSE [\ion{O}{II}] observations, such that we can be confident in these detections even if the error is significant. We only allow a minor offset in the velocity range where the stacked signal is assessed (the vertical dashed lines in Fig.~\ref{fig:stacks} are not exactly centered on zero).
Weighting the different spectra by their RMS noise instead of assuming equal weights lead to stacked integrated CO(3-2) line fluxes of $0.159~\rm Jy~km/s$ ($\rm SNR=2.9$) for all sources, $0.219~\rm Jy~km/s$ ($\rm SNR=2.8$) for the outflow cases, and a $0.220\rm~ Jy~km/s$ upper limit for the accretion cases.
Given the SNR, the relative uncertainty on the stacked CO line fluxes is $1/{\rm SNR}\sim30\%$, such that the values with and without weighting agree well with one another. 
Using the $r_{31}$ line ratio and the average redshifts, the stacked CO line fluxes and upper limits correspond respectively to intrinsic CO(1-0) luminosities equal to $2.0$, $2.9$, and $2.2~10^9~\rm K~km~s^{-1}~pc^2$. 
We further derive molecular gas masses, gas-to-stellar mass ratios, and depletion times as in Section~\ref{section:CO}, taking the average $\alpha_{\rm CO}$ conversion factor to estimate the molecular gas mass and the average SFR and stellar mass to estimate the mass ratio and depletion time. 
The $30\%$ uncertainty on the CO line fluxes together with the $30\%$ uncertainty on the Galactic conversion factor \citep{Bolatto2013}, the systematic difference up to 20\% between the metallicity corrections of \cite{Bolatto2013} and \cite{Genzel2012}, reflecting the scatter in the $\alpha_{\rm CO}-\rm metallicity$ relation, and a $20\%$ uncertainty on the line ratios \citep[e.g.,][]{Daddi2015} leads to a systematic uncertainty of at least 50\% (0.3 dex) on the stacked molecular gas masses. 

We include the stacked data points and upper limit in Figs.~\ref{fig:SFR_vs_LCO10}, \ref{fig:redshift} and \ref{fig:dMS}. The vertical error bars of the stacked data points assume 0.2 dex uncertainties in SFR and $M_\star$ in addition to the 0.3 dex uncertainty in molecular gas mass. The horizontal error bars reflect the $x$-axis spread of the data points. 
In Fig.~\ref{fig:redshift}, we can see that the average molecular gas-to-stellar mass ratio is slighly below its expected MS value given the redshifts and stellar masses of the sample due to the average $-0.23~\rm dex$ $\delta{\rm MS}$ offset below the MS, but that the average depletion time is well aligned with the expected MS value. 
In Fig.~\ref{fig:dMS}, we can see that both the stacked values for the whole sample and those for the outflow cases lie on their expected values from the \cite{Tacconi2018} scaling relations. The stacked upper limits for the accretion cases is also compatible with the scaling relations, notably as 65\% of the Gaussian distribution following the scatter in the scaling relations would fall below this upper limit. 
This indicates (i) that the molecular gas reservoirs of galaxies selected for their strong \ion{Mg}{II} absorption along quasar line-of-sights do not deviate significantly from those of mass-selected samples on average and (ii) that galaxies identified with accreting gas may not have specifically high molecular gas content. 
Point (i) notably means that the \cite{Tacconi2018} scaling relations can be used to estimate the expected CO fluxes from the MUSE [\ion{O}{II}] SFRs in future molecular gas follow-up studies of MEGAFLOW sources, as was already indicated in Fig.~\ref{fig:SFR_vs_LCO10}.
Point (ii) is potentially at odds with expectations from the quasi equilibrium and compaction models, since accretion should replenish the gas reservoirs and enhance star formation.
But the error bars of our stacked measurements are large, and mass-selected samples do include accreting and outflowing galaxies such that their molecular gas properties may already reflect the different configurations.
This trend needs to be confirmed by individual detections and larger samples, for example using the Atacama Large Millimeter/Submillimiter Array (ALMA).

\subsection{Dust continuum}

%3.3 Dust continuum upper limits: CO/LIR relation, comparison between the IR SFR upper limits with the [OII] SFRs -- maybe the SFRs are lower than anticipated? Maybe the galaxies are not very dusty and the CO lines weaker than expected? 

In addition to CO, the full bandwidth of the NOEMA observations enables to assess the dust continuum fluxes, indicated with their RMS errors in Table~\ref{table:continuum}. We report no individual detection, but these estimates enable to place individual 3$\sigma$ upper limits and can also be combined to provide a stacked result of $0.4 \pm 24~\rm \mu Jy$ as well as a $3\sigma$ stacked upper limit of $73~\rm \mu Jy$ at an effective frequency of 220 GHz, weighting each value by the RMS error. 

Long-wavelength dust emission can be used to probe the interstellar medium mass since the Rayleigh-Jeans tail of this emission is almost always optically thin \citep[e.g., ][]{Scoville2014,Scoville2016} and hence to provide an additional measurement of the gas mass. We estimate upper upper limits for the individual and stacked molecular gas masses from the dust continuum emission according to the \cite{Scoville2016} calibration, assuming a dust temperature $T_d=25~\rm K$ and an emissivity power-law index $\beta=1.8$. Namely, the molecular gas mass derived from the dust continuum is expressed as
\be
M_{\rm gas, dust}= \frac{s_{\rm dust} D_L^2}{\kappa(\nu_{\rm rest}) 2 kT_d (\nu_{\rm rest}/c)^2 \Gamma_{\rm RJ}(T_d, \nu_{\rm obs},z) (1+z)}
\ee
where $s_{\rm dust}$ is the observed flux density, $D_L$ the luminosity distance, $\kappa(\nu)=\kappa(\nu_{\rm 850\mu m}) (\lambda/{\rm 850\mu m})^{-\beta}$ the dust opacity, with $\kappa(\nu_{\rm 850\mu m})=4.84 ~10^{-4}~\rm  m^2~kg^{-1}$, and $\Gamma_{\rm RJ}(T_d, \nu_{\rm obs},z) = h\nu_{\rm obs}(1+z)/kT_d/(\exp(h\nu_{\rm obs}(1+z)/kT_d)-1)$ the correction for departure in the rest-frame  of the Planck function from Rayleigh-Jeans. 
The dust continuum emission can also translate into an infrared (IR) luminosity or equivalently in an IR SFR, $\rm SFR_{\rm IR}$, using the \cite{Magdis2012} templates with the mean redshift of the sample ($z=0.83$), the typical temperature for MS galaxies at this epoch and a \cite{Chabrier2003} IMF. 
We derive individual and stacked upper limits for these quantities, namely $M_{\rm gas, dust}<6.9~10^9~M_\odot$ and ${\rm SFR_{IR}}<15.2~M_\odot ~{\rm yr}^{-1}$ for the stack over all galaxies. 
Table~\ref{table:continuum} indicates the quantities derived from the dust continuum upper limits. 
We stress that a given observation can not simultaneously constrain $M_{\rm gas, dust}$ and $\rm SFR_{IR}$ since these quantities are not independent from each other here. 

\begin{table}
	\caption{Dust continuum measurements: RMS error, effective frequency, ISM mass, and IR SFR. The table includes the individual $3\sigma$ upper limits and the stacked values for all galaxies, for the accretion cases (A), and for the wind cases (W). We stress that the ISM mass and the IR SFR are not independent since they are both derived from the continuum flux. Galaxies are referred to with their short ID (cf. Table~\ref{fig:sample}).}
	\centering
	\begin{tabular}{lcccc}
		\hline
		\hline
		\noalign{\vskip 1mm}
		 ID &  $\sigma_{\rm cont}$ & $\nu_{\rm cont}$ & $M_{\rm gas, dust}$ &  $\rm SFR_{\rm IR}$\\
		   & [$\mu$Jy] & [GHz] &  [$M_\odot$] & $\rm [M_\odot ~yr^{-1}]$\\
		\hline
		\noalign{\vskip 1mm}
		0608 &     $55$   &    $209.1$&<$1.7~10^{10}$&<$24.3$ \\
		0912 &     $51$   &    $234.8$&<$1.2~10^{10}$&<$39.5$\\
		0949 &     $68$   &    $230.3$&<$1.7~10^{10}$&<$56.6$\\
		\hline
		\noalign{\vskip 1mm}
		1099 &     $75$   &    $213.5$&<$2.5~10^{10}$&<$72.0$\\
		0702 &     $45$   &    $209.0$&<$1.5~10^{10}$&<$23.8$\\
		0731&     $107$  &    $260.8$&<$1.5~10^{10}$&<$58.8$\\
		\hline
		\noalign{\vskip 1mm}
		stack & $24$ & $220.8$ &<$6.9~10^9$&<$15.2$\\
		A stack &$33$& $224.6$&<$8.7~10^9$&<$20.7$\\
		W stack &$36$& $216.1$&<$1.1~10^{10}$&<$22.3$\\
		\hline
	\end{tabular}
	\label{table:continuum}
\end{table}

The dust continuum molecular gas mass upper limits are consistent with the CO molecular gas upper limits indicated in Table~\ref{table:CO_derived} and fall within 0.3 dex of one another, which is the typical uncertainty of such measurements. The [\ion{O}{II}] SFR fall within the IR SFR upper limits. In particular, the $\rm SFR_{IR}$ stacked upper limit (<$15.2~\rm M_\odot yr^{-1}$) fits well with the average [\ion{O}{II}] SFR ($9.6 ~M_\odot~\rm yr^{-1}$).

\section{Conclusion}
\label{section:conclusion}

In this paper, we present the NOEMA MEGAFLOW pilot program, a first attempt at probing the molecular gas reservoirs of star-forming galaxies (SFGs) whose circum-galactic medium has been observed in absorption along quasar lines-of-sight and which thus have estimates of their mass accretion or ouflow rates. The motivation for such a program is to test the quasi equilibrium (`bathtub') and compaction models describing the evolution of star-forming galaxies along the main sequence (MS) of star formation, with simultaneous measurements of gas flows and gas content. We observe the CO(3-2) or CO(4-3) carbon monoxide line and the dust continuum of six galaxies at $z=0.6-1.1$ drawn from the MusE GAs FLOw and Wind (MEGAFLOW) survey \citep{Schroetter2016, Schroetter2019, Zabl2019, Zabl2020} with the IRAM NOEMA interferometer. Three of these galaxies were identified with accretion around the disc plane, three with outflows perpendicular to the disc (Table~\ref{table:sample} and Fig.~\ref{fig:sample}).
We report CO and dust continuum upper limits for the individual galaxies (Tables~\ref{table:CO_derived} and \ref{table:continuum}), and CO stacked detections over the whole sample and over the outflow cases (Fig.~\ref{fig:stacks}). From the CO data, we derive molecular gas masses, gas-to-stellar mass ratios and depletion times. From the dust continuum data, we derive upper limits for another estimate of the molecular gas content and the IR star formation rate. The dust continuum upper limits are compatible with the CO molecular gas mass upper limits and the SFR obtained from MUSE [\ion{O}{II}] observations. 
We further detect the dust continuum in three of the quasars and a strong line in one of them, which we identify with CO(4-3) (Appendix~\ref{appendix:quasar}).

Both the upper limits and the stacked results are compatible with the \cite{Tacconi2018} scaling relations obtained from a large sample of mass-selected SFGs on and around the MS (Figs.~\ref{fig:SFR_vs_LCO10}, \ref{fig:redshift} and \ref{fig:dMS}). This indicates that the molecular gas reservoirs of galaxies selected through their \ion{Mg}{II} absorption as the MEGAFLOW galaxies do not deviate significantly from mass-selected samples, such that the scaling relations can be used to estimate the expected CO fluxes from the Muse [\ion{O}{II}] SFRs in future studies. 
Galaxies identified with accretion may not have specifically high molecular gas content. This may be in tension with the theoretical expectations of the quasi equilibrium and compaction models, since accretion is associated with gas replenishment, but the low signal-to-noise ratios ($\rm SNR=3.2$ and $3.0$) of the stacked detections induce relatively large error bars. 

This pilot program establishes the feasability of molecular gas studies in galaxies with gas flows, such as those of the MEGAFLOW sample, and shows that molecular gas properties in absorption-selected samples may not fundamentally differ from those of mass-selected samples on average. This not only calls for additional observations to confirm or infirm the result of the stacked detections, but also provides tools to assess the observing time in future proposals, using the \cite{Tacconi2018} scaling relations. The next step consists in individual molecular gas detections in a statistically meaningful sample of galaxies with gas flows in order to quantify how the molecular gas fraction and depletion time depend on the inflow/outflow rates and test whether the gas content follows the expectation from the quasi equilibrium model and the compaction scenario. The current pilot program shows that while limited samples can be observed with NOEMA, ALMA is necessary to significantly increase the sample size and the sensitivity.

%%%%%%%%%%%%%%%%%%%%%%%%%%%%%%%%%%%%%%%%%%%%%%%%%%

\section*{Acknowledgements}

%The Acknowledgements section is not numbered. Here you can thank helpful colleagues, acknowledge funding agencies, telescopes and facilities used etc. Try to keep it short.

J. F. would like to thank F. Combes, P. Salom\'e, and A. Dekel for their valuable insights and advice, C. Lefèvre for the guidance during the data reduction, D. Maoz for his support. 
We thank the anonymous referee for his/her careful reading, which contributed to improve this article.
This work is based on observations carried out under project number W18CS with the IRAM NOEMA Interferometer. IRAM is supported by INSU/CNRS (France), MPG (Germany) and IGN (Spain).
This work was supported by the ANR 3DGasFlows (ANR-17-CE31-0017) and the OCEVU Labex (ANR-11-LABX-0060).
It has received funding from the European Research Council (ERC) under the European Union's FP7 Programme, Grant No. 833031, and was partly supported by the grants France-Israel PICS,
I-CORE Program of the
PBC/ISF 1829/12, BSF 2014-273, NSF AST-1405962,
GIF I-1341-303.7/2016, and DIP STE1869/2-1 GE625/17-1.

%%%%%%%%%%%%%%%%%%%%%%%%%%%%%%%%%%%%%%%%%%%%%%%%%%
\section*{Data Availability}

The data underlying this article will be shared on reasonable request to the corresponding author. 



\bibliographystyle{mnras}
\bibliography{freundlich_megaflow} % if your bibtex file is called example.bib

%\begin{thebibliography}{99}
%\end{thebibliography}

%%%%%%%%%%%%%%%%%%%%%%%%%%%%%%%%%%%%%%%%%%%%%%%%%%

%%%%%%%%%%%%%%%%% APPENDICES %%%%%%%%%%%%%%%%%%%%%

\appendix

\section{CO spectra}

\begin{figure*}
	\centering
	\includegraphics[width=0.48\linewidth]{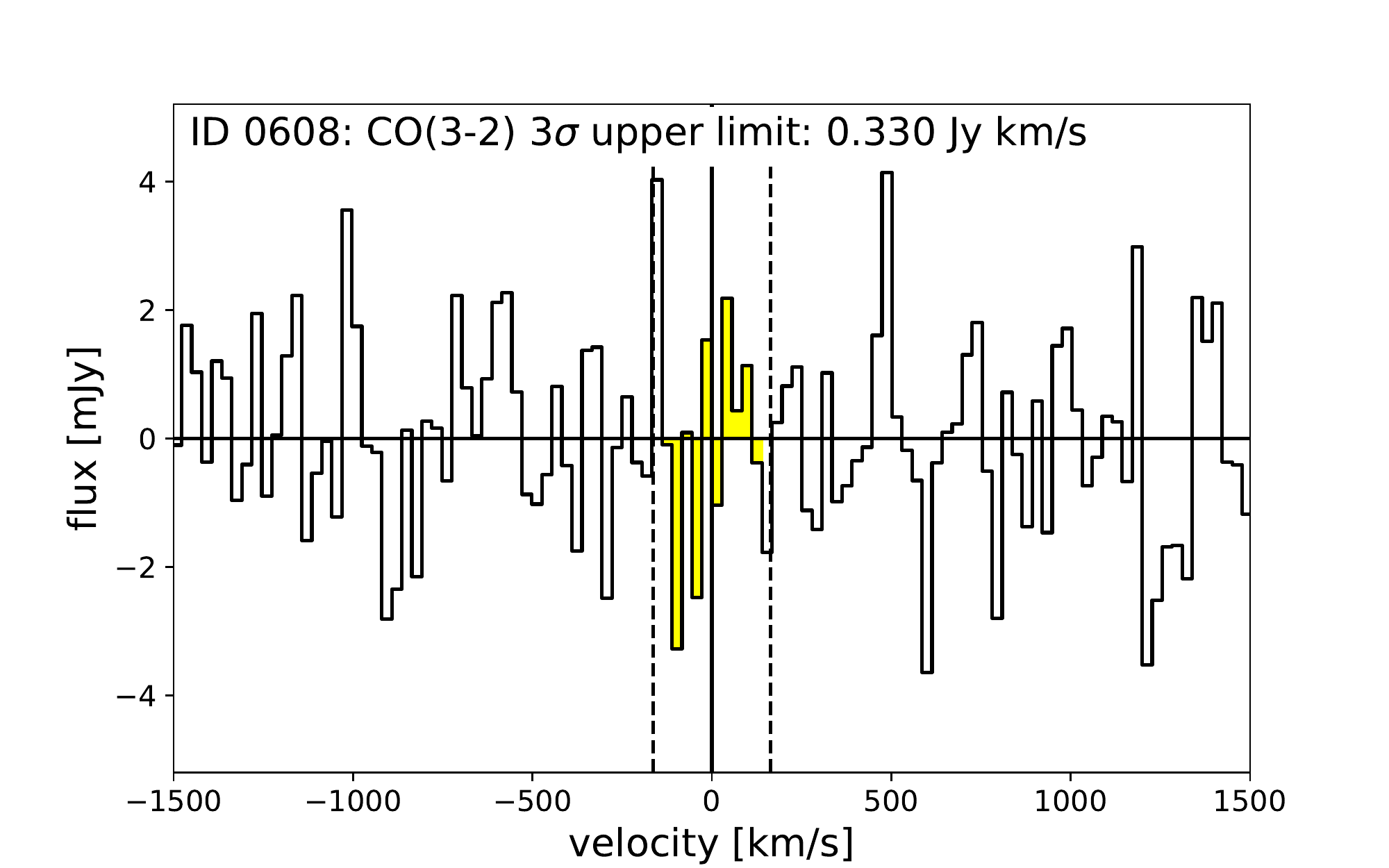}
	\includegraphics[width=0.48\linewidth]{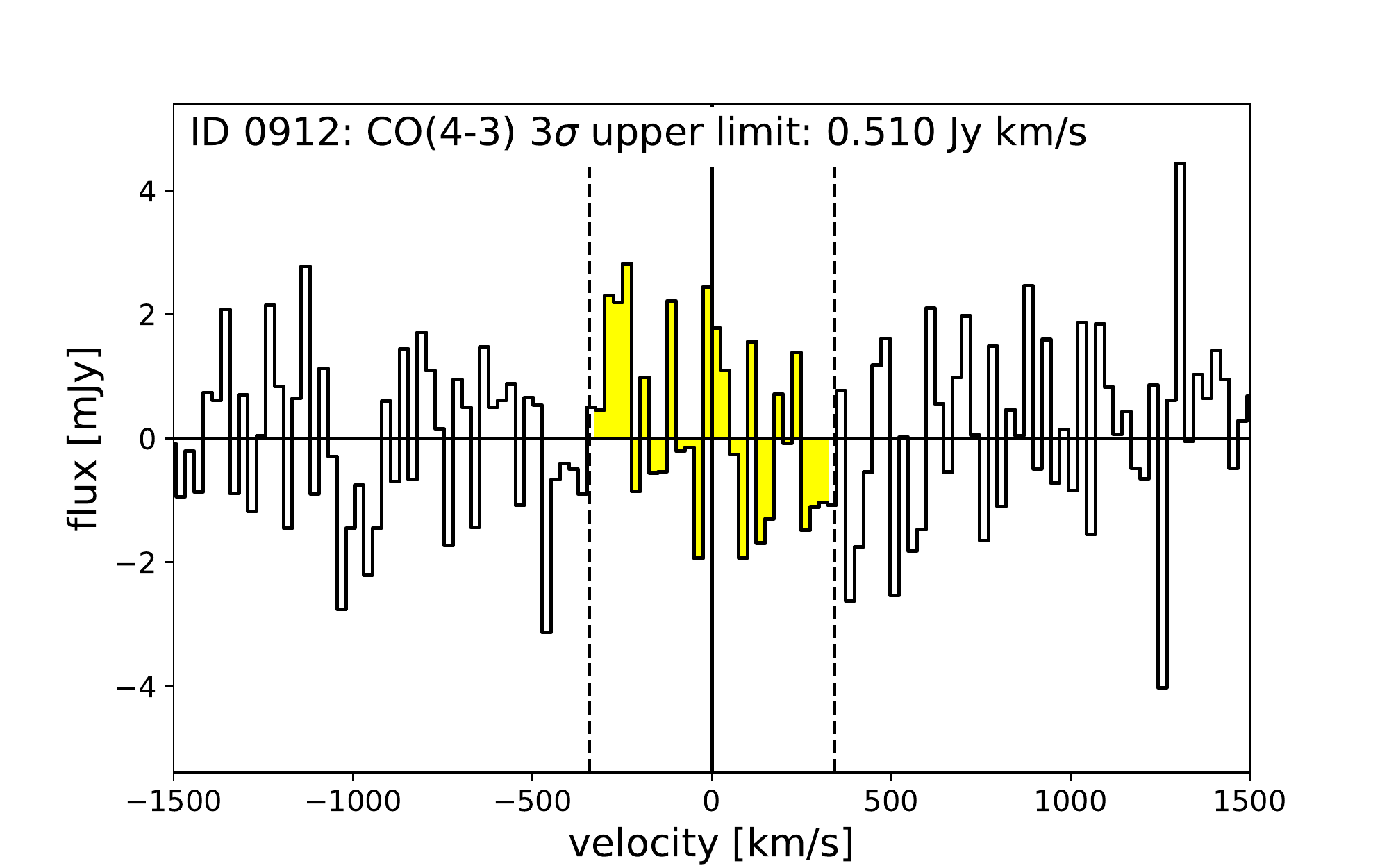}
	\includegraphics[width=0.48\linewidth]{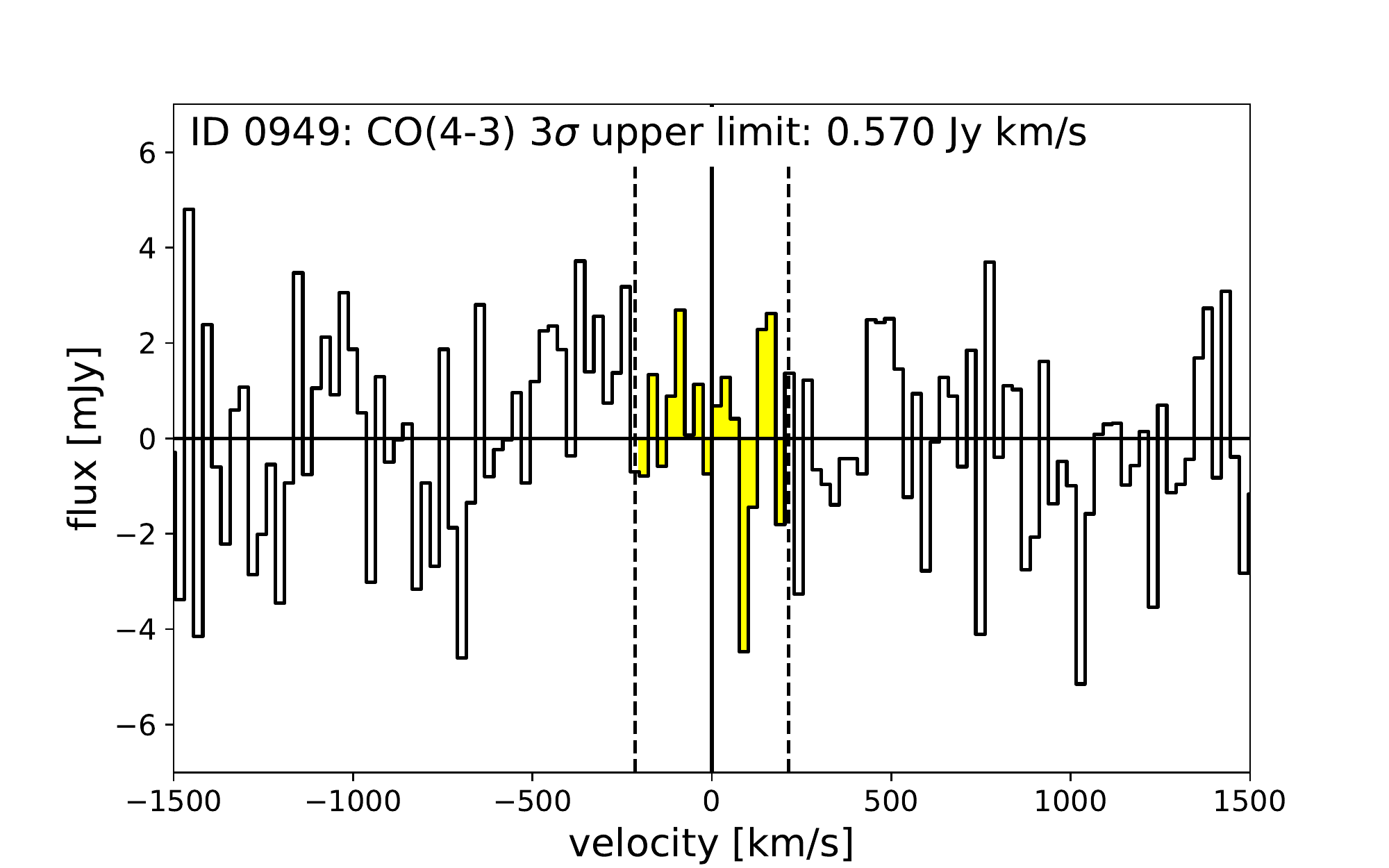}
	\includegraphics[width=0.48\linewidth]{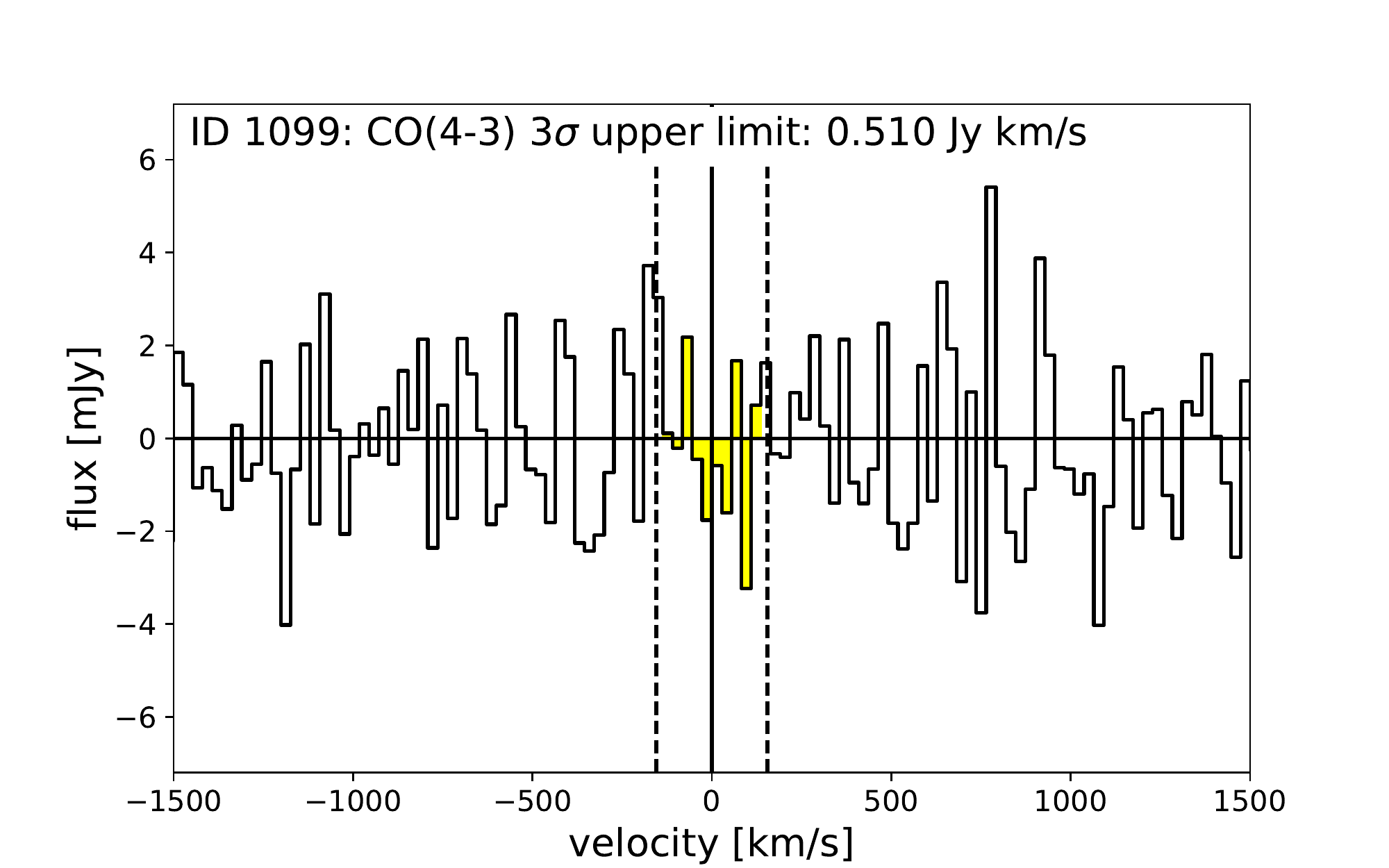}
	\includegraphics[width=0.48\linewidth]{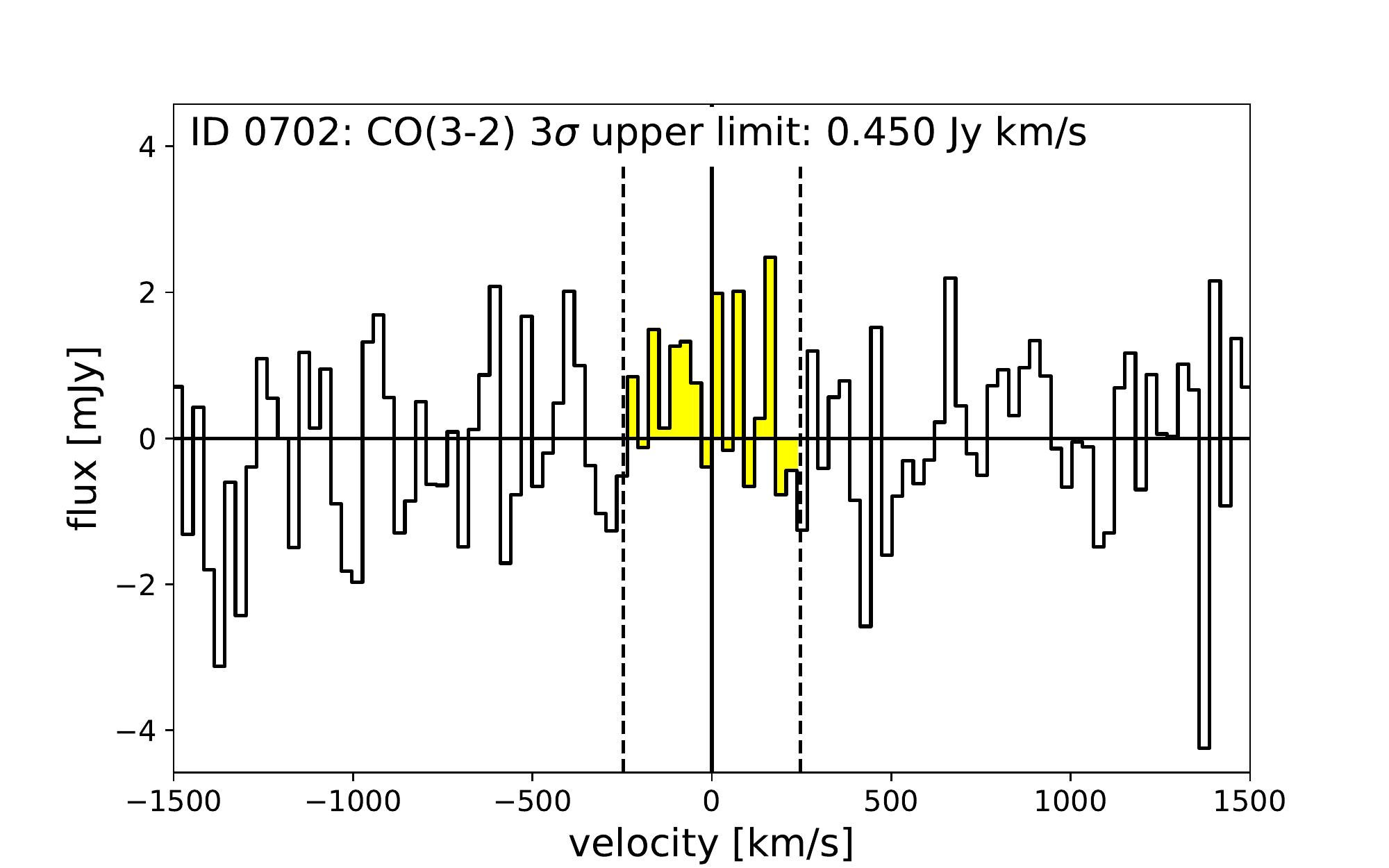}
	\includegraphics[width=0.48\linewidth]{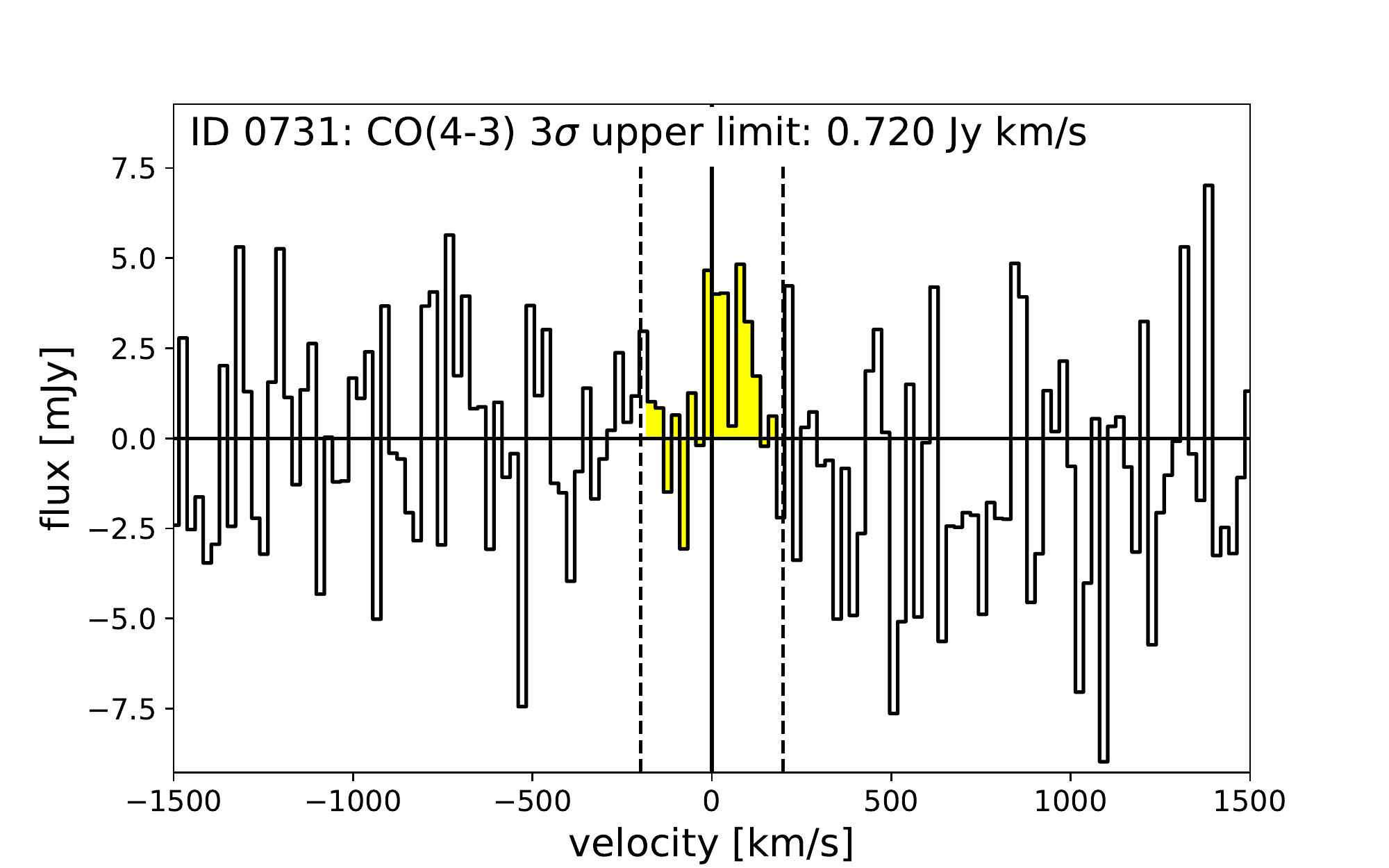}
	\caption{CO spectra of the six galaxies of the NOEMA MEGAFLOW sample extracted at the optical position, fitting the PSF in $uv$ space. The plain vertical line corresponds to $v=0\rm ~km/s$, the two vertical dashed lines highlight the FWTM derived from the MUSE [OII] FWHM assuming a Gaussian line. 
	We calculate the signal within the FWTM but do not find any detection with $\rm SNR \geq 3$. Galaxies  0702 and 0731 may display hints of detection at low SNR,  with integrated line fluxes within the FWTM respectively equal to $\rm 0.280 ~Jy~km/s$ ($\rm SNR=1.9$) and $\rm 0.567~Jy~km/s$ ($\rm SNR=2.4$). Galaxies are referred to with their short ID (cf. Table~\ref{fig:sample}).}
	\label{fig:spectra}
\end{figure*}

We display in Fig.~\ref{fig:spectra} the CO spectra for the six  galaxies of the sample observed with NOEMA. 

\section{Quasar detections}
\label{appendix:quasar}

\begin{figure*}
	\centering
	\includegraphics[width=0.32\linewidth,trim={2.cm 0.cm 2.8cm 0cm},clip]{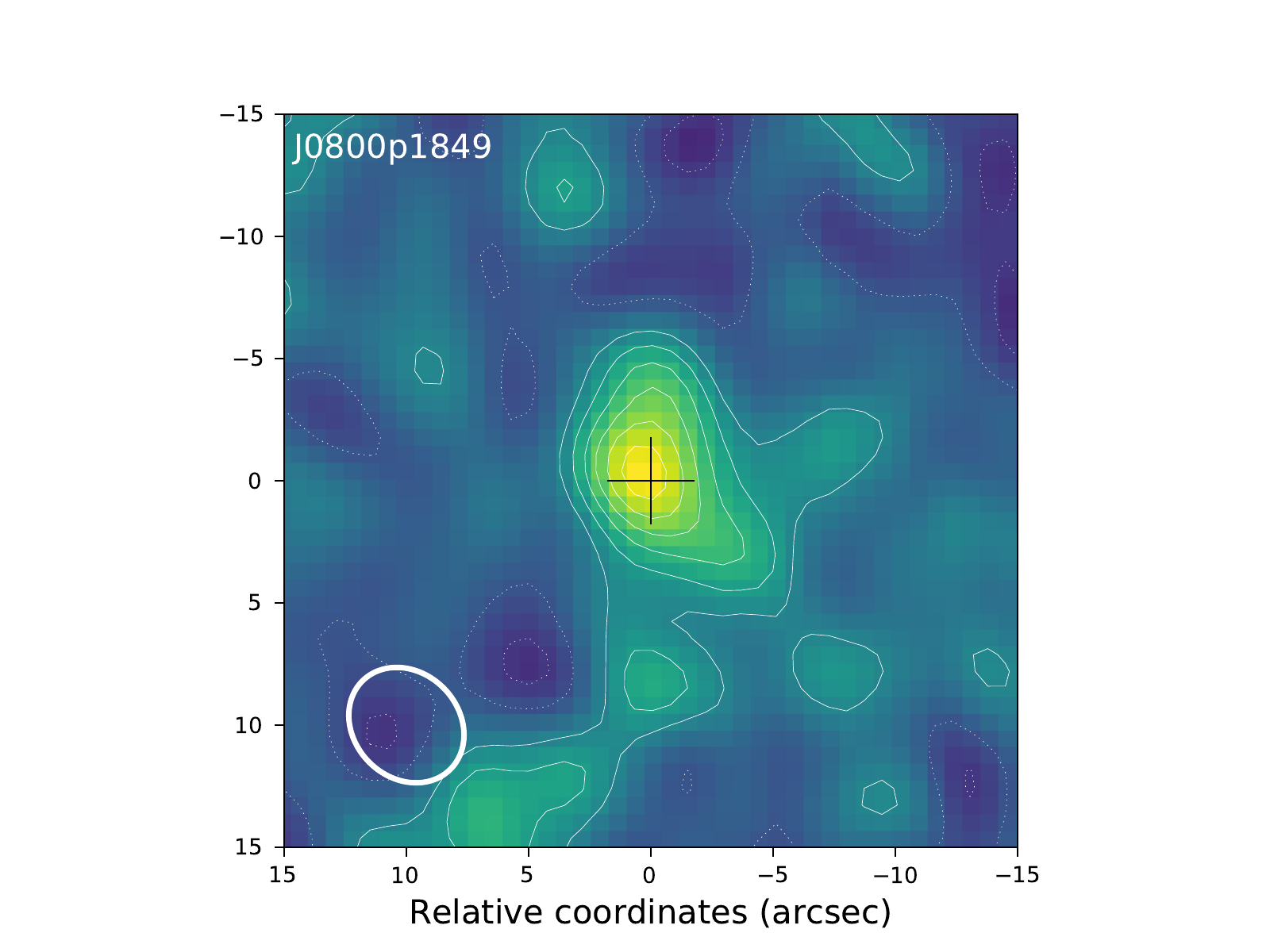}
	\hfill
	\includegraphics[width=0.32\linewidth,trim={2.cm 0.cm 2.8cm 0cm},clip]{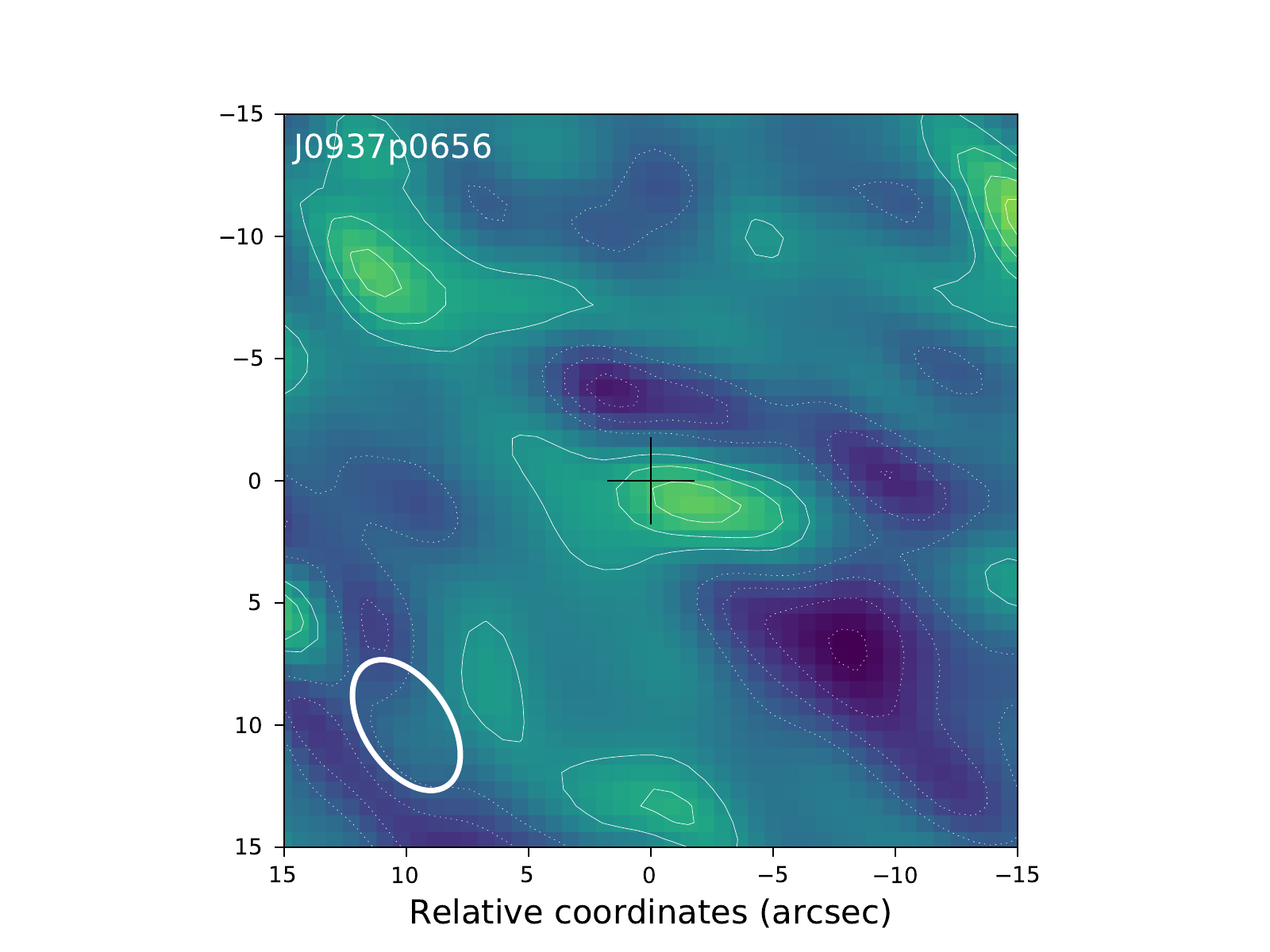}
	\hfill
	\includegraphics[width=0.32\linewidth,trim={2.cm 0.cm 2.8cm 0cm},clip]{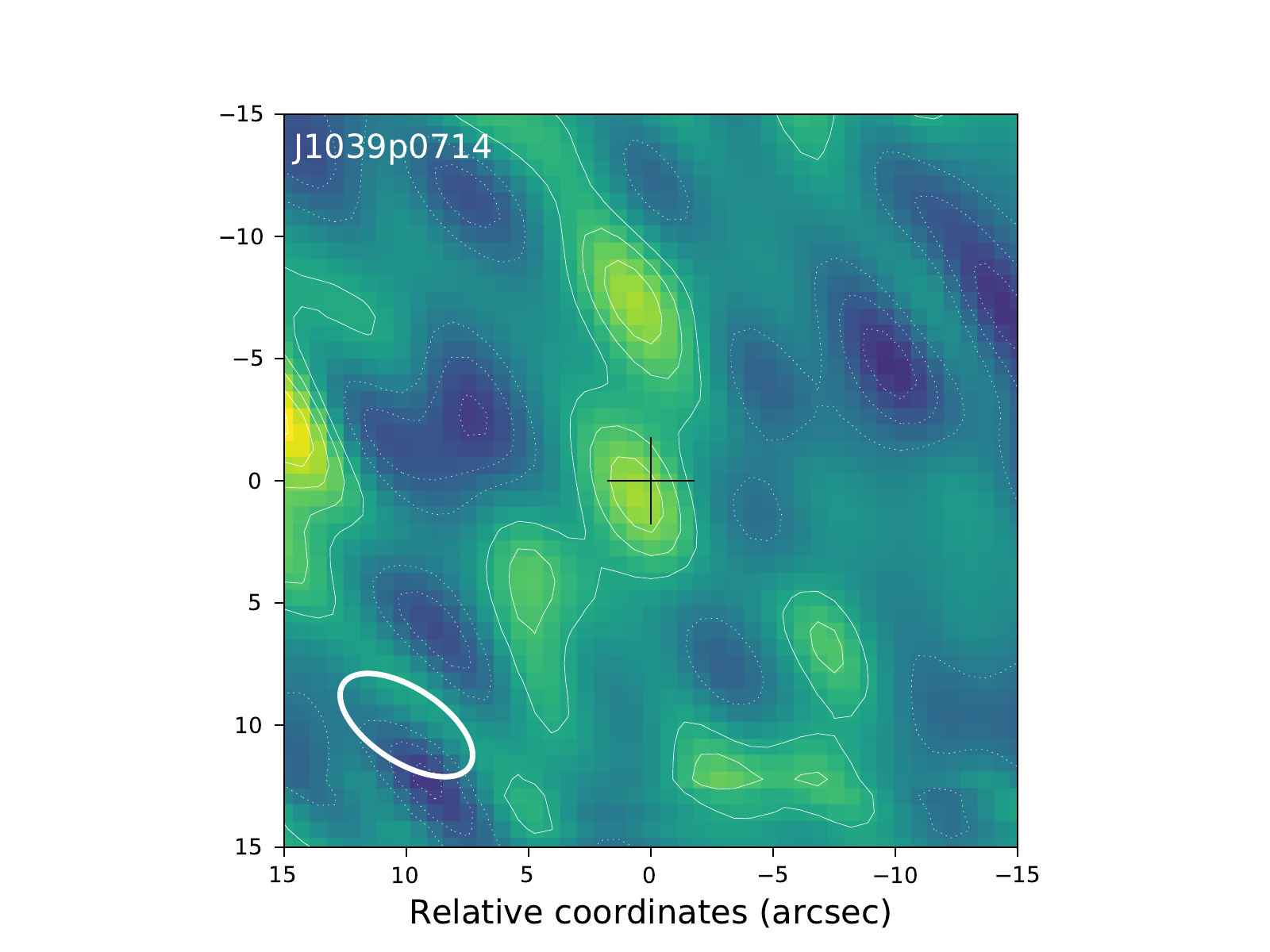}
	\caption{Continuum maps of the detected quasars J0800p1849, J0937p0656, and J1039p0714. The effective frequencies are 209.1, 208.6, and 230.3 GHz. The central cross highlights the quasar location from the \protect\cite{Zhu2013} catalog. Contours correspond to integer $\sigma$ levels (dotted for the negative ones). The white ellipse at the bottom left indicates the beam.}
	\label{fig:qso_maps}
\end{figure*}

Although our main targets are the galaxies observed in MgII absorption, we clearly detect the dust continuum emission of quasars J0800p1849 and J0937p0656 and tentatively that of J1039p0714. Upper limits can be determined from the RMS noise for the other three quasars. Table~\ref{table:quasars} summarizes these continuum observations, while Fig.~\ref{fig:qso_maps} shows the continuum maps associated to the detected quasars.
Quasar J0800p1849 is spatially resolved with a FWHM of $1.65 \pm 0.22~\rm arcsec$ (14 kpc at the quasar redshift, $z=1.294$), which may indicate a large host galaxy.

\begin{figure}
	\centering
	\includegraphics[width=1\linewidth,trim={.6cm 0.cm 1.5cm 0.5cm},clip]{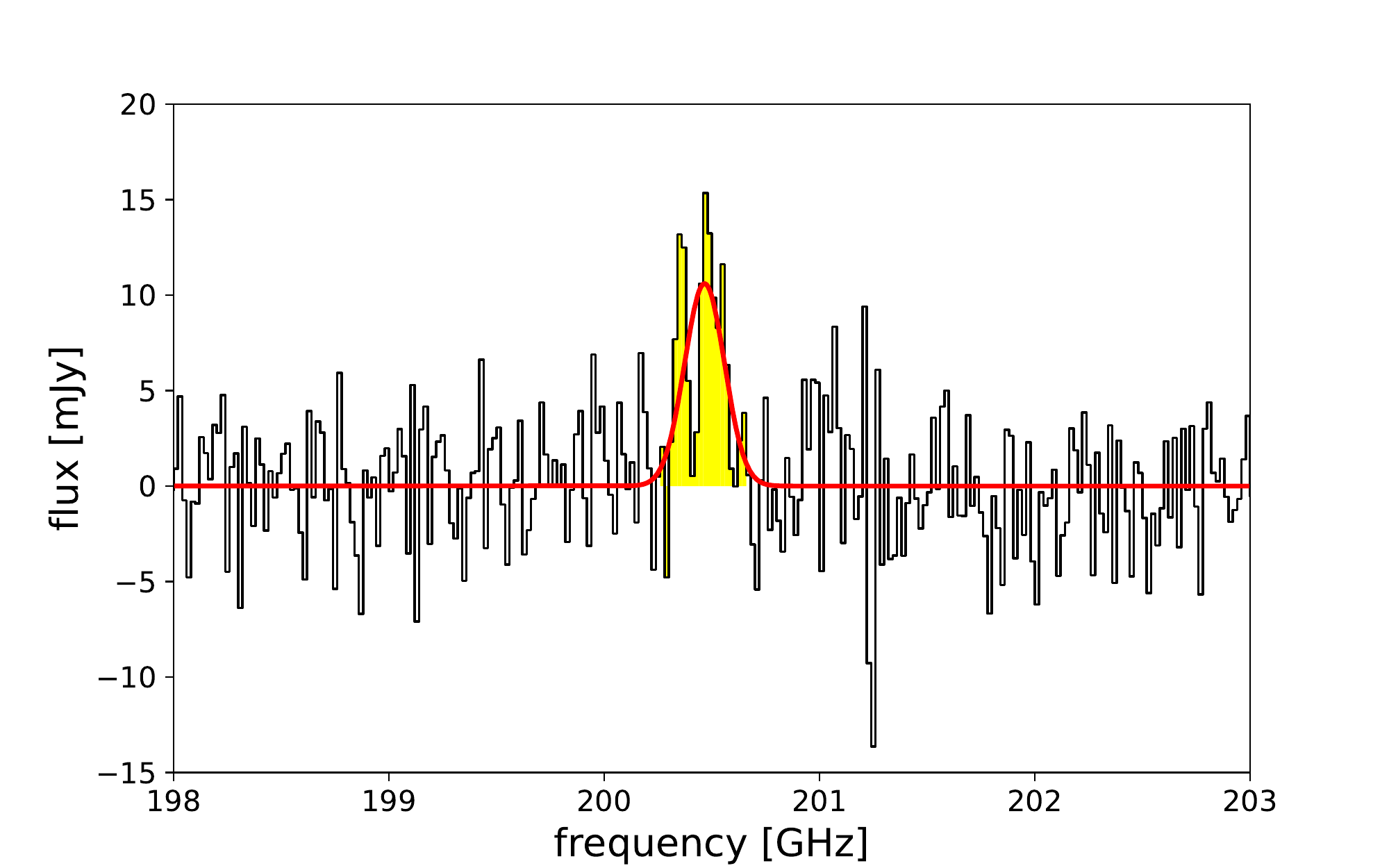}
	\vspace{-0.2cm}
	\caption{Spectrum of quasar J0800p1849, displaying a bright emission line. We carried out a Gaussian fit to the line using \texttt{gildas class}, yielding $\nu_{\rm obs} = 200.47 \pm 0.02\rm~GHz$, a line width equal to $\rm 326 \pm 48~km/s$, and an integrated line flux $3.7\pm0.5~\rm Jy~km/s$. }
	\label{fig:qso_maser}
\end{figure}

We further extract the spectrum at the optical position of each quasar by fitting the PSF in the $uv$ space with \texttt{go uvfit}. Although most of these spectra do not display any specific feature, we observe a bright SNR=7.3 emission line at $\nu_{\rm obs} = 200.466\rm~GHz$ for J0800p1849. The spectrum is shown in Fig.~\ref{fig:qso_maser}, and we fit the line with a Gaussian to assess its integrated flux. 
This line may be associated with CO(4-3) ($\nu_{\rm rest}=461.041~\rm GHz$) at the quasar redshift ($z=1.294$), albeit with a $\sim700~\rm km/s$ velocity offset which could be due to broad features, wings, or uncertainties in the quasar redshift. This association is reinforced by the presence of a line that could correspond to [\ion{O}{II}] with a comparable offset in the MUSE data. Assuming a Galactic conversion factor and a \mbox{CO(4-3)/CO(1-0)} ratio of 0.42, the integrated line flux yields a molecular gas mass of $2.2~10^{11}M_\odot$. Alternatively, the line could also correspond to the emission of an otherwise undetected galaxy on the line of sight, for example CO(2-1) ($\nu_{\rm rest}=230.538~\rm GHz$) at $z=0.150$, since none of the detected \ion{Mg}{II} absorbers on the line-of-sight of this quasar is expected to display strong emission lines around $\nu_{\rm obs} = 200.466\rm~GHz$.

\begin{table*}
	\caption{Quasar continuum observations: ID, coordinates and redshift, continuum effective frequency, RMS noise, detected flux or $3\sigma$ upper limit, and SNR.}
	%. (1) Galaxy ID. (2)(3)(4)(5) Quasar identifier, R.A., DEC. coordinates, and redshift, as listed in \protect\cite{Zabl2019}. (6)(7) Continuum flux and RMS noise. (8) Continuum effective frequency. }
	\label{table:quasars}
	\centering
	\begin{tabular}{llllllll}
	\hline
	\hline
	\noalign{\vskip 1mm}
	Quasar ID & R.A. & DEC. & $z$ & $\nu_{\rm cont}~\rm [GHz]$ & $\sigma_{\rm cont}~\rm [mJy]$ & $s_{\rm cont}~\rm [mJy]$  & SNR \\
	\noalign{\vskip 1mm}
	%(1) & (2)  & (3) & (4) & (5)  & (6) & (7) & (8)\\
	\hline
	\noalign{\vskip 1mm}
	J0800p1849 &   08:00:04.55265 &+18:49:35.0828  &$1.294$&209.1&0.107&0.485&4.5$^\star$\\
	J1236p0725 &   12:36:24.3931  &+07:25:51.5496  &$1.606$&234.9&0.051&<$0.153$    &-\\
	 J1039p0714 &   10:39:36.66840 & +07:14:27.36   &$1.536$&230.3&0.077 &0.179 &2.3\\
	 J0838p0257 &   08:38:52.05480 &+02:57:03.657   &$1.771$&213.7&0.088&<$0.263$     &-\\
	 J0937p0656 &   09:37:49.5866  &+06:56:56.2704  &$1.802$&208.6&0.053&0.354&6.7\\
	J0015m0751 &   00:15:35.17573 &-07:51:03.07357 &$0.874$&260.4&0.111&<$0.334$     &-\\
	\hline
	\end{tabular}
	\vspace{0.2cm}
	\\
	\begin{minipage}{1\linewidth}
	%\justify
	\textbf{Notes.} We use the optical position for the quasar, except for J0800p1849 and J0937p065 where the signal is strong enough to release this prior. Typical astrometric corrections are at the level of 0.1-0.2 arcsec. $^\star$The SNR associated to J0800p1849 is artificially low compared to the other two because of the increased number of degrees of freedom enabled by its spatial resolution: the source is indeed resolved (cf. Fig.~\ref{fig:qso_maps}), so we fit it with a Gaussian rather than a point source. The FWHM of the Gaussian is found to be $1.65 \pm 0.22
    ~\rm arcsec$, i.e., about 14 kpc at the quasar redshift.
    \end{minipage}
\end{table*}

% Don't change these lines
\bsp	% typesetting comment
\label{lastpage}
\end{document}